\documentclass[showpacs,preprintnumbers,twocolumn,superscriptaddress,aps,nofootinbib]{revtex4-1}
\usepackage{amssymb}
\usepackage[centertags]{amsmath}
\usepackage{txfonts}
\usepackage{epsfig}
\usepackage{bm}
\usepackage{color}
\usepackage{graphicx,graphics}
\usepackage{multirow}
\usepackage{float}
\usepackage{ulem}
\usepackage{appendix}
\usepackage[pdfstartview=FitH]{hyperref}
\hypersetup{colorlinks=true, citecolor=blue,linkcolor=blue,filecolor=black,urlcolor=blue}  % comment when submit
\allowdisplaybreaks[2]

\begin{document}

\title{Different coalescence sources of light nuclei production in Au-Au collisions at $\sqrt{s_{NN}}=3$ GeV}

\author{Rui-Qin Wang}
%\email {wangrq@qfnu.edu.cn}
\affiliation{School of Physics and Physical Engineering, Qufu Normal University, Shandong 273165, China}

\author{Ji-Peng L\"{u}}
\affiliation{School of Physics and Physical Engineering, Qufu Normal University, Shandong 273165, China}

\author{Yan-Hao Li}
\affiliation{School of Physics and Physical Engineering, Qufu Normal University, Shandong 273165, China}

\author{Jun Song}
%\email {songjun2011@jnxy.edu.cn}
\affiliation{School of Physical Science and Intelligent Engineering, Jining University, Shandong 273155, China}

\author{Feng-Lan Shao}
\email {shaofl@mail.sdu.edu.cn}
\affiliation{School of Physics and Physical Engineering, Qufu Normal University, Shandong 273165, China}

\begin{abstract}

We study the production of light nuclei in the coalescence mechanism in Au-Au collisions at midrapidity at $\sqrt{s_{NN}}=3$ GeV. 
We derive analytic formulas of momentum distributions of two bodies, three bodies and four nucleons coalescing into light nuclei, respectively.
We naturally explain the transverse momentum spectra of the deuteron ($d$), triton ($t$), helium-3 ($^3$He) and helium-4 ($^4$He). 
We reproduce the data of yield rapidity densities and averaged transverse momenta of $d$, $t$, $^3$He and $^4$He.
We give proportions of contributions from different coalescence sources for $t$, $^3$He and $^4$He in their productions.
We find that besides nucleon coalescence, nucleon$+$nucleus coalescence and nucleus$+$nucleus coalescence may play requisite roles in light nuclei production in Au-Au collisions at $\sqrt{s_{NN}}=3$ GeV.

\end{abstract}

\maketitle

%-------------------------------------------------------------------------------------------------- Introduction
\section{Introduction}

As a specific group of observables in relativistic heavy ion collisions~\cite{Nagle:1994hm,Chen:2018tnh,Blum:2019suo,Bazak:2020wjn,Aichelin:1991xy,Gutbrod:1976zzr,Junnarkar:2019equ,Andronic:2017pug,Bzdak:2019pkr,Sun:2017xrx,Sun:2018jhg,Luo:2020pef}, 
light nuclei such as the deuteron ($d$), triton ($t$), helium-3 ($^3$He) and helium-4 ($^4$He) have always been under active investigation in recent decades
both in experiment~\cite{STAR:2001pbk,PHENIX:2007tef,NA49:2016qvu,Albergo:2002gi,ALICE:2015wav,ALICE:2020chv,STAR:2016ydv,STAR:2020hya,Zhang:2020ewj,STAR:2019sjh,STAR:2022hbp} and in theory~\cite{Braun-Munzinger:2018hat,Oliinychenko:2020ply,Dover:1991zn,Chen:2003qj,Mrowczynski:2020ugu,Andronic:2010qu}.
The STAR experiment at the BNL Relativistic Heavy Ion Collider (RHIC) and the ALICE experiment at the CERN Large Hadron Collider (LHC) have collected a wealth of data on light nuclei production.
These data exhibit some fascinating features, especially their non-trivial energy-dependent behaviors in a wide collision energy range from GeV to TeV magnitude~\cite{ALICE:2015wav,ALICE:2020chv,STAR:2016ydv,STAR:2020hya,Zhang:2020ewj,STAR:2019sjh,STAR:2022hbp}.
Theoretical studies have also made significant progress.
Two production mechanisms, the thermal production mechanism~\cite{Mekjian:1977ei,Siemens:1979dz,Andronic:2010qu,Cleymans:2011pe,Cai:2019jtk} and the coalescence mechanism~\cite{Schwarzschild:1963zz,Sato:1981ez,Dover:1991zn,Mattiello:1995xg,Nagle:1996vp,Mattiello:1996gq,Chen:2003qj,Polleri:1997bp,Scheibl:1998tk,Sharma:2018dyb,Bazak:2018hgl}, have proved to be successful in describing light nuclei formation. 
In addition, transport scenario~\cite{Danielewicz:1991dh,Oh:2009gx,Oliinychenko:2018ugs,Oliinychenko:2020znl,Staudenmaier:2021lrg,Kireyeu:2022qmv} is employed to study how light nuclei evolve and survive during the hadronic system evolution.

The coalescence mechanism, in which light nuclei are usually assumed to be produced by the coalescence of the jacent nucleons in the phase space, possesses its unique characteristics.
Plenty of current experimental observations at high RHIC and LHC energies favor the nucleon coalescence~\cite{STAR:2016ydv,STAR:2019sjh,STAR:2022hbp,ALICE:2020chv,Zhao:2018lyf,Zhao:2020irc,Wang:2020zaw}.
Recently the STAR collaboration has extended the beam energy scan program to lower collision energy and published the data of both hadrons and light nuclei in Au-Au collisions at $\sqrt{s_{NN}}=3$ GeV~\cite{STAR:2021yiu,STAR:2021ozh,STAR:2021hyx,Liu:2021xfy}.
These data show very different properties compared to those at high RHIC and LHC energies, such as the disappearance of partonic collectivity~\cite{STAR:2021yiu} and dominant baryonic interactions~\cite{STAR:2021ozh}.
At this low collision energy besides nucleons, light nuclei in particular of light $d$, $t$ and $^3$He have been more abundantly created~\cite{Liu:2021xfy} compared to higher collision energies~\cite{Bartsch:2021ozc}.
It is easier in physics for these light nuclei to capture nucleons or other light nuclei  to form heavier composite objects.
In fact clear depletions below unity of proton$-d$ and $d-d$ correlation functions measured at such low collision energy indicate the strong final state interaction and further support the possible coalescence of the $d$ with the nucleon or other $d$~\cite{Mi:2022zig}.
How much space is there on earth for other particle coalescence except nucleons, e.g., composite particles of less mass numbers coalescing into light nuclei of larger mass numbers or composite particles capturing nucleons to recombine into heavier light nuclei?

In this article, we extend the coalescence model which has been successfully used to explain the momentum dependence of yields and coalescence factors of different light nuclei at high RHIC and LHC energies~\cite{Wang:2020zaw,Zhao:2022xkz}, to include nucleon$+$nucleus coalescence and nucleus$+$nucleus coalescence besides nucleon coalescence.
We apply the extended coalescence model to hadronic systems created in Au-Au collisions at midrapidity area at $\sqrt{s_{NN}}=3$ GeV to study the momentum and centrality dependence of light nuclei production in the low- and intermediate-$p_T$ regions.
We compute the transverse momentum ($p_T$) spectra, the yield rapidity densities ($dN/dy$) and the averaged transverse momenta ($\langle p_T \rangle$) of $d$, $t$, $^3$He and $^4$He from central to peripheral collisions.
We give proportions of contributions from different coalescence sources for $t$, $^3$He and $^4$He respectively in their productions.
Our studies show that in $0-10$\%, $10-20$\% and $20-40$\% centralities, besides nucleon coalescence, nucleon$+d$ coalescence plays an important role in $t$ and $^3$He production
and nucleon$+d$ ($t$, $^3$He) coalescence as well as $d+d$ coalescence occupy significant proportions in $^4$He production.
But in the peripheral $40-80$\% centrality, nucleon coalescence plays a dominant role, and nucleon$+$nucleus coalescence or nucleus$+$nucleus coalescence seems to disappear.

The rest of the paper is organized as follows. 
In Sec.~\ref{model}, we introduce the coalescence model. 
We present analytic formulas of momentum distributions of two bodies, three bodies, and four nucleons coalescing into light nuclei, respectively.
In Sec.~\ref{results}, we apply the model to Au-Au collisions in different rapidity intervals at midrapidity area at $\sqrt{s_{NN}}=3$ GeV to study momentum and centrality dependence of the production of various species of light nuclei in the low- and intermediate-$p_T$ regions.
We give proportions of contributions from different coalescence sources for $t$, $^3$He and $^4$He in their productions.
In Sec.~\ref{summary} we summarize our work.

%----------------------------------------------------------------------------------------------------------------------------------------------------------------- Model
\section{The coalescence model}   \label{model}

In this section we introduce the coalescence model which is used to deal with the light nuclei production.
The starting point of the model is a hadronic system produced at the late stage of the evolution of high energy collision.
The hadronic system consists of different species of primordial mesons and baryons.
In the first step of the model all primordial nucleons are allowed to form $d$, $t$, $^3$He and $^4$He via the nucleon coalescence.
Then in the second step the formed $d$, $t$ and $^3$He capture the remanent primordial nucleons, i.e., those excluding consumed ones in the nucleon coalescence process, or other light nuclei to recombine into nuclei with larger mass numbers.
In this model only $d$, $t$, $^3$He and $^4$He are included, and those light nuclei with mass number larger than 4 are abandoned.

In the following we present the deduction of the formalism of the production of various species of light nuclei via different coalescence processes, respectively. 
First we give analytic results of two bodies coalescing into light nuclei, which can be applied to processes such as $p+n \rightarrow d$, $n+d \rightarrow t$, 
$p+d \rightarrow ^3$He, $p+t \rightarrow ^4$He, $n+^3$He $\rightarrow ^4$He and $d+d \rightarrow ^4$He.
Then we show analytic results of three bodies coalescing into light nuclei, which can be used to describe these processes, e.g., $n+n+p \rightarrow t$, $p+p+n \rightarrow ^3$He and $p+n+d \rightarrow ^4$He.
Finally, we give the analytic result of four nucleons coalescing into $^4$He, i.e., $p+p+n+n \rightarrow ^4$He. 

%--------------------------------------------------------------------------------------------------  h1h2 --> Lj2h
\subsection{Formalism of two bodies coalescing into light nuclei}      \label{2hcoLj}

%-------------------------------------------------------------------------------------------------- general formalism
We begin with a hadronic system produced at the final stage of the evolution of high energy collision and suppose light nuclei $L_{j}$ are formed via the coalescence of two hadronic bodies $h_1$ and $h_2$.
The three-dimensional momentum distribution of the produced light nuclei $f_{L_{j}}(\bm{p})$ is given by
{\setlength\arraycolsep{0pt}
\begin{eqnarray}
 f_{L_{j}}(\bm{p}) =&&  \int d\bm{x}_1d\bm{x}_2 d\bm{p}_1 d\bm{p}_2  f_{h_1h_2}(\bm{x}_1,\bm{x}_2;\bm{p}_1,\bm{p}_2)  \nonumber  \\
 &&~~ \times \mathcal {R}_{L_{j}}(\bm{x}_1,\bm{x}_2;\bm{p}_1,\bm{p}_2,\bm{p}),      \label{eq:fLj2hgeneral} 
\end{eqnarray} }%
where $f_{h_1h_2}(\bm{x}_1,\bm{x}_2;\bm{p}_1,\bm{p}_2)$ is two-hadron joint coordinate-momentum distribution; 
$\mathcal {R}_{L_{j}}(\bm{x}_1,\bm{x}_2;\bm{p}_1,\bm{p}_2,\bm{p})$ is the kernel function.
Here and from now on we use bold symbols to denote three-dimensional coordinate or momentum vectors.

In terms of the normalized joint coordinate-momentum distribution denoted by the superscript `$(n)$', we have
{\setlength\arraycolsep{0pt}
\begin{eqnarray}
 f_{L_{j}}(\bm{p})=&& N_{h_1h_2} \int d\bm{x}_1d\bm{x}_2 d\bm{p}_1 d\bm{p}_2  f^{(n)}_{h_1h_2}(\bm{x}_1,\bm{x}_2;\bm{p}_1,\bm{p}_2)  \nonumber \\
  && ~~~~~~~~~~~ \times \mathcal {R}_{L_{j}}(\bm{x}_1,\bm{x}_2;\bm{p}_1,\bm{p}_2,\bm{p}).      \label{eq:fLj2hgeneral1}               
\end{eqnarray} }%
$N_{h_1h_2}$ is the number of all possible $h_1h_2$-pairs, and it is equal to $N_{h_1}N_{h_2}$ and $N_{h_1}(N_{h_1}-1)$ for $h_1 \neq h_2$ and $h_1=h_2$, respectively.
$N_{h_i}~ (i=1,2)$ is the number of the hadrons $h_i$ in the considered hadronic system.

The kernel function $\mathcal {R}_{L_{j}}(\bm{x}_1,\bm{x}_2;\bm{p}_1,\bm{p}_2,\bm{p})$ denotes the probability density for $h_1$, $h_2$ with
momenta $\bm{p}_1$ and $\bm{p}_2$ at $\bm{x}_1$ and $\bm{x}_2$ to recombine into a $L_{j}$ of momentum $\bm{p}$.
It carries the kinetic and dynamical information of $h_1$ and $h_2$ recombining into light nuclei,
and its precise expression should be constrained by such as the momentum conservation, constraints due to intrinsic quantum numbers e.g. spin, and so on~\cite{Wang:2017vsm,Wang:2020zaw,Zhao:2022xkz}.
To take these constraints into account explicitly, we rewrite the kernel function in the following form
{\setlength\arraycolsep{0pt}
\begin{eqnarray}
  \mathcal {R}_{L_{j}}(\bm{x}_1,\bm{x}_2;\bm{p}_1,\bm{p}_2,\bm{p}) =&& g_{L_{j}} \mathcal {R}_{L_{j}}^{(x,p)}(\bm{x}_1,\bm{x}_2;\bm{p}_1,\bm{p}_2) \nonumber \\  
 && \times \delta(\displaystyle{\sum^2_{i=1}} \bm{p}_i-\bm{p}),     \label{eq:RLj2h}  
\end{eqnarray} }%
where the spin degeneracy factor $g_{L_{j}} = (2J_{L_{j}}+1) /[\prod \limits_{i=1}^2(2J_{h_i}+1)]$.
$J_{L_{j}}$ is the spin of the produced $L_{j}$ and $J_{h_i}$ is that of the primordial hadron $h_i$. 
The Dirac $\delta$ function guarantees the momentum conservation in the coalescence.
The remaining $\mathcal {R}_{L_{j}}^{(x,p)}(\bm{x}_1,\bm{x}_2;\bm{p}_1,\bm{p}_2)$ can be solved from the Wigner transformation once the wave function of $L_{j}$ is given with the instantaneous coalescence approximation.
It is as follows 
{\setlength\arraycolsep{0pt}
\begin{eqnarray}
&&  \mathcal {R}^{(x,p)}_{L_{j}}(\bm{x}_1,\bm{x}_2;\bm{p}_1,\bm{p}_2) = 8e^{-\frac{(\bm{x}'_1-\bm{x}'_2)^2}{2\sigma^2}}
     e^{-\frac{2\sigma^2(m_2\bm{p}'_{1}-m_1\bm{p}'_{2})^2}{(m_1+m_2)^2\hbar^2c^2}},      \label{eq:RLj2hxp} 
\end{eqnarray} }%
as we adopt the wave function of a spherical harmonic oscillator as in Refs.~\cite{Chen:2003ava,Zhu:2015voa}.
The superscript `$'$' in the coordinate or momentum variable denotes the hadronic coordinate or momentum in the rest frame of the $h_1h_2$-pair.
$m_1$ and $m_2$ are the rest mass of hadron $h_1$ and that of hadron $h_2$.
The width parameter $\sigma=\sqrt{\frac{2(m_1+m_2)^2}{3(m_1^2+m_2^2)}} R_{L_{j}}$,
where $R_{L_{j}}$ is the root-mean-square radius of $L_{j}$ and its values for different light nuclei can be found in Ref.~\cite{Angeli:2013epw}.
The factor $\hbar c$ comes from the used GeV$\cdot$fm unit, and it is 0.197 GeV$\cdot$fm.

%-------------------------------------------------------------------------------------------------- factorization assumption 
The normalized two-hadron joint distribution $f^{(n)}_{h_1h_2}(\bm{x}_1,\bm{x}_2;\bm{p}_1,\bm{p}_2)$ is generally coordinate and momentum coupled, 
especially in central heavy-ion collisions with relatively high collision energies where the collective expansion exists long. 
The coupling intensities and its specific forms are probably different at different phase spaces in different collision energies and different collision centralities. 
In this article, we try our best to derive production formulas analytically and present centrality and momentum dependence of light nuclei more intuitively in Au-Au collisions at low RHIC energy $\sqrt{s_{NN}}=3$ GeV where the partonic collectivity disappears~\cite{STAR:2021yiu}, 
so we consider a simple case that the joint distribution is coordinate and momentum factorized, i.e.,
{\setlength\arraycolsep{0pt}
\begin{eqnarray}
&& f^{(n)}_{h_1h_2}(\bm{x}_1,\bm{x}_2;\bm{p}_1,\bm{p}_2) = f^{(n)}_{h_1h_2}(\bm{x}_1,\bm{x}_2)   f^{(n)}_{h_1h_2}(\bm{p}_1,\bm{p}_2).  \label{eq:fh1h2fac}   
\end{eqnarray} }%

Substituting Equations~(\ref{eq:RLj2h}-\ref{eq:fh1h2fac}) into Equation~(\ref{eq:fLj2hgeneral1}), we have
{\setlength\arraycolsep{0.2pt}
\begin{eqnarray}
&& f_{L_{j}}(\bm{p})= N_{h_1h_2} g_{L_{j}} \int d\bm{x}_1d\bm{x}_2 f^{(n)}_{h_1h_2}(\bm{x}_1,\bm{x}_2)  8e^{-\frac{(\bm{x}'_1-\bm{x}'_2)^2}{2\sigma^2}}  \nonumber   \\
&&~~~~~~ \times
 \int d\bm{p}_1d\bm{p}_2 f^{(n)}_{h_1h_2}(\bm{p}_1,\bm{p}_2)  e^{-\frac{2\sigma^2(m_2\bm{p}'_{1}-m_1\bm{p}'_{2})^2}{(m_1+m_2)^2\hbar^2c^2}}
   \delta(\displaystyle{\sum^2_{i=1}} \bm{p}_i-\bm{p})   \nonumber   \\
&&~~~~~~~~~ = N_{h_1h_2} g_{L_{j}} \mathcal {A}_{L_{j}}  \mathcal {M}_{L_{j}}(\bm{p}),    \label{eq:fLj2h}  
\end{eqnarray} }%
where we use $\mathcal {A}_{L_{j}}$ to denote the coordinate integral part in Equation~(\ref{eq:fLj2h}) as
{\setlength\arraycolsep{0pt}
\begin{eqnarray}
\mathcal {A}_{L_{j}} =  8\int d\bm{x}_1d\bm{x}_2 f^{(n)}_{h_1h_2}(\bm{x}_1,\bm{x}_2) e^{-\frac{(\bm{x}'_1-\bm{x}'_2)^2}{2\sigma^2}},      \label{eq:ALj2h}      
\end{eqnarray} }%
and use $\mathcal {M}_{L_{j}}(\bm{p})$ to denote the momentum integral part as
{\setlength\arraycolsep{0pt}
\begin{eqnarray}
 \mathcal {M}_{L_{j}}(\bm{p}) = 
    \int d\bm{p}_1d\bm{p}_2 f^{(n)}_{h_1h_2}(\bm{p}_1,\bm{p}_2)  e^{-\frac{2\sigma^2(m_2\bm{p}'_{1}-m_1\bm{p}'_{2})^2}{(m_1+m_2)^2\hbar^2c^2}}
   \delta(\displaystyle{\sum^2_{i=1}} \bm{p}_i-\bm{p}).  \nonumber \\   \label{eq:MLj2h}
\end{eqnarray} }%
$\mathcal {A}_{L_{j}}$ stands for the probability of a $h_1h_2$-pair satisfying the coordinate requirement to recombine into $L_j$,
and $\mathcal {M}_{L_{j}}(\bm{p})$ stands for the probability density of a $h_1h_2$-pair satisfying the momentum requirement to recombine into $L_j$ with momentum $\bm{p}$.

%------------------------------------------------------------------------------------------------------------- A_{Lj2h}  
Changing integral variables in Equation~(\ref{eq:ALj2h}) to be $\bm{X}= \frac{\bm{x}_1+\bm{x}_2}{\sqrt{2}}$ and $\bm{r}= \frac{ \bm{x}_1-\bm{x}_2}{\sqrt{2}}$, we have
{\setlength\arraycolsep{0pt}
\begin{eqnarray}
 \mathcal {A}_{L_{j}} =  8\int d\bm{X} d\bm{r} f^{(n)}_{h_1h_2}(\bm{X},\bm{r}) e^{-\frac{\bm{r}'^2}{\sigma^2}},   \label{eq:ALj2hXr}   
\end{eqnarray} }%
and the normalizing condition
{\setlength\arraycolsep{0pt}
\begin{eqnarray}
 \int f^{(n)}_{h_1h_2}(\bm{X},\bm{r})  d\bm{X}d\bm{r}=1.
\end{eqnarray} }%
We further assume the coordinate joint distribution is coordinate variable factorized, i.e., $f^{(n)}_{h_1h_2}(\bm{X},\bm{r}) = f^{(n)}_{h_1h_2}(\bm{X}) f^{(n)}_{h_1h_2}(\bm{r})$.
Adopting $f^{(n)}_{h_1h_2}(\bm{r}) = \frac{1}{(\pi C_w R_f^2)^{3/2}} e^{-\frac{\bm{r}^2}{C_w R_f^2}}$ as in Refs.~\cite{Mrowczynski:2016xqm,Wang:2020zaw}, we have
{\setlength\arraycolsep{0pt}
\begin{eqnarray}
&& \mathcal {A}_{L_{j}} =  \frac{8}{(\pi C_w R_f^2)^{3/2}} \int d\bm{r} e^{-\frac{\bm{r}^2}{C_wR_f^2}} e^{-\frac{\bm{r}'^2}{\sigma^2}}.   \label{eq:ALj2h-r}  
\end{eqnarray} }%
Here $R_f$ is the effective radius of the hadronic system at the light nuclei freeze-out.
$C_w$ is a distribution width parameter and it is set to be 2, the same as that in Refs.~\cite{Mrowczynski:2016xqm,Wang:2020zaw}. 

Considering instantaneous coalescence in the rest frame of $h_1h_2$-pair, i.e., $\Delta t'=0$, we get
\begin{eqnarray}
\bm{r} = \bm{r}' +(\gamma-1)\frac{\bm{r}'\cdot \bm{\beta}}{\beta^2}\bm{\beta},    \label{eq:LorentzTr}  
\end{eqnarray}
where $\bm{\beta}$ is the three-dimensional velocity vector of the center-of-mass frame of $h_1h_2$-pair in the laboratory frame and the Lorentz contraction factor $\gamma=1/\sqrt{1-\bm{\beta}^2}$.
Substituting Equation~(\ref{eq:LorentzTr}) into Equation~(\ref{eq:ALj2h-r}) and integrating from the relative coordinate variable, we can obtain 
{\setlength\arraycolsep{0pt}
\begin{eqnarray}
 \mathcal {A}_{L_{j}} = \frac{8\sigma^3}{(C_w R_f^2+\sigma^2) \sqrt{C_w (R_f/\gamma)^2+\sigma^2}}.   \label{eq:ALj2h-final}  
\end{eqnarray} }%

%------------------------------------------------------------------------------------------------------------- M_{Lj2h} 
Noticing that $\hbar c/\sigma$ in Equation~(\ref{eq:MLj2h}) has a small value of about 0.1, we can mathematically approximate the gaussian form of the momentum-dependent kernel function to be a $\delta$ function form as follows
\begin{equation}
e^{-\frac{(\bm{p}'_{1}-\frac{m_1}{m_2}\bm{p}'_{2})^2} {(1+\frac{m_1}{m_2})^2 \frac{\hbar^2c^2}{2\sigma^2}}} \approx
\left[ \frac{\hbar c}{\sigma} (1+\frac{m_1}{m_2}) \sqrt{\frac{\pi}{2}} \right]^3 \delta(\bm{p}'_{1}-\frac{m_1}{m_2}\bm{p}'_{2}).
\end{equation}
After integrating $\bm{p}_1$ and $\bm{p}_2$ from Equation~(\ref{eq:MLj2h}) we can obtain
{\setlength\arraycolsep{0.2pt}
\begin{eqnarray}
 \mathcal {M}_{L_{j}}(\bm{p}) =  (\frac{\hbar c\sqrt{\pi}}{\sqrt{2}\sigma})^3 \gamma  f^{(n)}_{h_1h_2}(\frac{m_1\bm{p}}{m_1+m_2},\frac{m_2\bm{p}}{m_1+m_2}),  \label{eq:MLj2h-pfin}
\end{eqnarray} }%
where $\gamma$ comes from $\bm{p}'_{1}-\frac{m_1}{m_2}\bm{p}'_{2}=\frac{1}{\gamma} (\bm{p}_{1}-\frac{m_1}{m_2}\bm{p}_{2})$.

%------------------------------------------------------------------------------------------------------------- f_{Lj2h}
Substituting Equations~(\ref{eq:ALj2h-final}) and (\ref{eq:MLj2h-pfin}) into Equation~(\ref{eq:fLj2h}) and ignoring correlations between $h_1$ and $h_2$ hadrons, we have
{\setlength\arraycolsep{0.2pt}
\begin{eqnarray}
 f_{L_{j}}(\bm{p})  &=& \frac{ (\sqrt{2\pi}\hbar c)^3 g_{L_{j}} \gamma}{(C_w R_f^2+\sigma^2) \sqrt{C_w (R_f/\gamma)^2+\sigma^2}} 
                       f_{h_1}(\frac{m_1\bm{p}}{m_1+m_2})     \nonumber   \\
  &&  \times f_{h_2}(\frac{m_2\bm{p}}{m_1+m_2}).    \label{eq:fLj2h-approx}
\end{eqnarray} }%
Denoting the Lorentz invariant momentum distribution $\dfrac{d^{2}N}{2\pi p_{T}dp_{T}dy}$ with $f^{(inv)}$, we finally have
{\setlength\arraycolsep{0.2pt}
\begin{eqnarray}
f_{L_j}^{(inv)}(p_{T},y) &=&\frac{ (\sqrt{2\pi}\hbar c)^3 g_{L_{j}} }{(C_w R_f^2+\sigma^2) \sqrt{C_w (R_f/\gamma)^2+\sigma^2}}  \frac{m_1+m_2}{m_1m_2}   \nonumber   \\
  &&  \times        f_{h_1}^{(inv)}(\frac{m_1p_{T}}{m_1+m_2},y)  f_{h_2}^{(inv)}(\frac{m_2p_{T}}{m_1+m_2},y) ,     \label{eq:pt-Lj2h}
\end{eqnarray} }%
where $y$ is the rapidity.

%--------------------------------------------------------------------------------------------------  h1h2h3 --> Lj3h
\subsection{Formalism of three bodies coalescing into light nuclei}

%-------------------------------------------------------------------------------------------------- general formalism
For light nuclei $L_{j}$ formed via the coalescence of three hadronic bodies $h_1$, $h_2$ and $h_3$, the three-dimensional momentum distribution $f_{L_{j}}(\bm{p})$ is
{\setlength\arraycolsep{0pt}
\begin{eqnarray}
 f_{L_{j}}(\bm{p})=&& N_{h_1h_2h_3} \int d\bm{x}_1d\bm{x}_2d\bm{x}_3 d\bm{p}_1 d\bm{p}_2d\bm{p}_3  f^{(n)}_{h_1h_2h_3}(\bm{x}_1,\bm{x}_2,\bm{x}_3;   \nonumber \\
  &&   \bm{p}_1,\bm{p}_2,\bm{p}_3)  \mathcal {R}_{L_{j}}(\bm{x}_1,\bm{x}_2,\bm{x}_3;\bm{p}_1,\bm{p}_2,\bm{p}_3,\bm{p}).     \label{eq:fLj3hgeneral1}               
\end{eqnarray} }%
$N_{h_1h_2h_3}$ is the number of all possible $h_1h_2h_3$-clusters and it is equal to $N_{h_1}N_{h_2}N_{h_3},~N_{h_1}(N_{h_1}-1)N_{h_3},~N_{h_1}(N_{h_1}-1)(N_{h_1}-2)$ for $h_1 \neq h_2 \neq h_3$, $h_1 = h_2 \neq h_3$, $h_1=h_2=h_3$, respectively.
$f^{(n)}_{h_1h_2h_3}$ is the normalized three-hadron joint coordinate-momentum distribution.
$\mathcal {R}_{L_{j}}$ is the kernel function.

We rewrite the kernel function as
{\setlength\arraycolsep{0pt}
\begin{eqnarray}
  \mathcal {R}_{L_{j}}(\bm{x}_1,\bm{x}_2,\bm{x}_3;\bm{p}_1,\bm{p}_2,\bm{p}_3,\bm{p}) =&& g_{L_{j}}
 \mathcal {R}_{L_{j}}^{(x,p)}(\bm{x}_1,\bm{x}_2,\bm{x}_3;\bm{p}_1,\bm{p}_2,\bm{p}_3) \nonumber \\  
 && \times \delta(\displaystyle{\sum^3_{i=1}} \bm{p}_i-\bm{p}).     \label{eq:RLj3h}  
\end{eqnarray} }%
The spin degeneracy factor $g_{L_{j}} = (2J_{L_{j}}+1) /[\prod \limits_{i=1}^3(2J_{h_i}+1)]$.
The Dirac $\delta$ function guarantees the momentum conservation.
$\mathcal {R}_{L_{j}}^{(x,p)}(\bm{x}_1,\bm{x}_2,\bm{x}_3;\bm{p}_1,\bm{p}_2,\bm{p}_3)$ solving from the Wigner transformation~\cite{Chen:2003ava,Zhu:2015voa} is
{\setlength\arraycolsep{0pt}
\begin{eqnarray}
 && \mathcal {R}^{(x,p)}_{L_{j}}(\bm{x}_1,\bm{x}_2,\bm{x}_3;\bm{p}_1,\bm{p}_2,\bm{p}_3) = 8^2 e^{-\frac{(\bm{x}'_1-\bm{x}'_2)^2}{2\sigma_1^2}}
      e^{-\frac{2(\frac{m_1\bm{x}'_1}{m_1+m_2}+\frac{m_2\bm{x}'_2}{m_1+m_2}-\bm{x}'_3)^2}{3\sigma_2^2}}   \nonumber   \\
  &&~~ \times  e^{-\frac{2\sigma_1^2(m_2\bm{p}'_{1}-m_1\bm{p}'_{2})^2}{(m_1+m_2)^2\hbar^2c^2}}
     e^{-\frac{3\sigma_2^2[m_3\bm{p}'_{1}+m_3\bm{p}'_{2}-(m_1+m_2)\bm{p}'_{3}]^2} {2(m_1+m_2+m_3)^2\hbar^2c^2}}.      \label{eq:RLj3hxp} 
\end{eqnarray} }%
The superscript `$'$' denotes the hadronic coordinate or momentum in the rest frame of the $h_1h_2h_3$-cluster.
The width parameter $\sigma_1=\sqrt{\frac{m_3(m_1+m_2)(m_1+m_2+m_3)} {m_1m_2(m_1+m_2)+m_2m_3(m_2+m_3)+m_3m_1(m_3+m_1)}} R_{L_{j}}$,
and $\sigma_2=\sqrt{\frac{4m_1m_2(m_1+m_2+m_3)^2} {3(m_1+m_2)[m_1m_2(m_1+m_2)+m_2m_3(m_2+m_3)+m_3m_1(m_3+m_1)]}} R_{L_{j}}$.

%-------------------------------------------------------------------------------------------------- factorization assumption 
With the coordinate and momentum factorization assumption of the joint distribution,
we have
{\setlength\arraycolsep{0.2pt}
\begin{eqnarray}
 f_{L_{j}}(\bm{p}) = N_{h_1h_2h_3} g_{L_{j}} \mathcal {A}_{L_{j}}  \mathcal {M}_{L_{j}}(\bm{p}).   \label{eq:fLj3h}  
\end{eqnarray} }%
Here we also use $\mathcal {A}_{L_{j}}$ to denote the coordinate integral part as
{\setlength\arraycolsep{0pt}
\begin{eqnarray}
\mathcal {A}_{L_{j}} = &&  8^2\int d\bm{x}_1d\bm{x}_2d\bm{x}_3 f^{(n)}_{h_1h_2h_3}(\bm{x}_1,\bm{x}_2,\bm{x}_3) e^{-\frac{(\bm{x}'_1-\bm{x}'_2)^2}{2\sigma_1^2}}   \nonumber  \\
&&~~~~~~~~ \times  e^{-\frac{2(\frac{m_1\bm{x}'_1}{m_1+m_2}+\frac{m_2\bm{x}'_2}{m_1+m_2}-\bm{x}'_3)^2}{3\sigma_2^2}} ,      \label{eq:ALj3h}      
\end{eqnarray} }%
and use $\mathcal {M}_{L_{j}}(\bm{p})$ to denote the momentum integral part as
{\setlength\arraycolsep{0pt}
\begin{eqnarray}
 \mathcal {M}_{L_{j}}(\bm{p}) = &&
    \int d\bm{p}_1d\bm{p}_2d\bm{p}_3 f^{(n)}_{h_1h_2h_3}(\bm{p}_1,\bm{p}_2,\bm{p}_3)   \delta(\displaystyle{\sum^3_{i=1}} \bm{p}_i-\bm{p})   \nonumber   \\
&&~~~~ \times    e^{-\frac{2\sigma_1^2(m_2\bm{p}'_{1}-m_1\bm{p}'_{2})^2}{(m_1+m_2)^2\hbar^2c^2}}
     e^{-\frac{3\sigma_2^2[m_3\bm{p}'_{1}+m_3\bm{p}'_{2}-(m_1+m_2)\bm{p}'_{3}]^2} {2(m_1+m_2+m_3)^2\hbar^2c^2}} .  \nonumber \\   \label{eq:MLj3h}
\end{eqnarray} }%

%------------------------------------------------------------------------------------------------------------- A_{Lj3h}
We change integral variables in Equation~(\ref{eq:ALj3h}) to be $\bm{Y}= (m_1\bm{x}_1+m_2\bm{x}_2+m_3\bm{x}_3)/(m_1+m_2+m_3)$, $\bm{r}_1= (\bm{x}_1-\bm{x}_2)/\sqrt{2}$ and 
$\bm{r}_2=\sqrt{\frac{2}{3}} (\frac{m_1\bm{x}_1}{m_1+m_2}+\frac{m_2\bm{x}_2}{m_1+m_2}-\bm{x}_3)$, 
and further assume the coordinate joint distribution is coordinate variable factorized, i.e., 
$3^{3/2} f^{(n)}_{h_1h_2h_3}(\bm{Y},\bm{r}_1,\bm{r}_2) = f^{(n)}_{h_1h_2h_3}(\bm{Y}) f^{(n)}_{h_1h_2h_3}(\bm{r}_1) f^{(n)}_{h_1h_2h_3}(\bm{r}_2)$.
Adopting $f^{(n)}_{h_1h_2h_3}(\bm{r}_1) = \frac{1}{(\pi C_1 R_f^2)^{3/2}} e^{-\frac{\bm{r}_1^2}{C_1 R_f^2}}$ 
and $f^{(n)}_{h_1h_2h_3}(\bm{r}_2) = \frac{1}{(\pi C_2 R_f^2)^{3/2}} e^{-\frac{\bm{r}_2^2}{C_2 R_f^2}}$ as in Refs.~\cite{Mrowczynski:2016xqm,Wang:2020zaw}, we have
{\setlength\arraycolsep{0pt}
\begin{eqnarray}
 \mathcal {A}_{L_{j}} &=& 8^2  \frac{1}{(\pi C_1 R_f^2)^{3/2}} \int d\bm{r}_1 e^{-\frac{\bm{r}_1^2}{C_1 R_f^2}} e^{-\frac{(\bm{r}'_1)^2}{\sigma_1^2}}  \nonumber  \\
   && \times         \frac{1}{(\pi C_2 R_f^2)^{3/2}} \int d\bm{r}_2 e^{-\frac{\bm{r}_2^2}{C_2 R_f^2}} e^{-\frac{(\bm{r}'_2)^2}{\sigma_2^2}}.   \label{eq:ALj3h-r}  
\end{eqnarray} }%
Comparing relations of $\bm{r}_1$, $\bm{r}_2$ with $\bm{x}_1$, $\bm{x}_2$, $\bm{x}_3$ to that of $\bm{r}$ with $\bm{x}_1$, $\bm{x}_2$ in Sec.~\ref{2hcoLj}, 
we see that $C_1$ is equal to $C_w$ and $C_2$ is $4C_w/3$ when ignoring the mass difference of $m_1$ and $m_2$~\cite{Mrowczynski:2016xqm,Wang:2020zaw}. 
Considering the Lorentz transformation and integrating from the relative coordinate variables in Equation~(\ref{eq:ALj3h-r}), we obtain 
{\setlength\arraycolsep{0pt}
\begin{eqnarray}
 \mathcal {A}_{L_{j}} &=&  \frac{8^2 \sigma_1^3\sigma_2^3} {  (C_1 R_f^2+\sigma_1^2) \sqrt{C_1 (R_f/\gamma)^2+\sigma_1^2} }  \nonumber  \\
  && \times \frac{1}{ (C_2 R_f^2+\sigma_2^2) \sqrt{C_2 (R_f/\gamma)^2+\sigma_2^2}} .   \label{eq:ALj3h-final}  
\end{eqnarray} }%

%------------------------------------------------------------------------------------------------------------- M_{Lj3h} 
Approximating the gaussian form of the momentum-dependent kernel function to be $\delta$ function form 
and integrating $\bm{p}_1$, $\bm{p}_2$ and $\bm{p}_3$ from Equation~(\ref{eq:MLj3h}), we can obtain
{\setlength\arraycolsep{0.2pt}
\begin{eqnarray}
 && \mathcal {M}_{L_{j}}(\bm{p}) =  \left( \frac{\pi \hbar^2 c^2}{\sqrt{3}\sigma_1\sigma_2}  \right)^3   \gamma^2 \times  \nonumber   \\
  && ~ f^{(n)}_{h_1h_2h_3}(\frac{m_1\bm{p}}{m_1+m_2+m_3},\frac{m_2\bm{p}}{m_1+m_2+m_3},\frac{m_3\bm{p}}{m_1+m_2+m_3}). ~~~~~ \label{eq:MLj3h-pfin}
\end{eqnarray} }%

%------------------------------------------------------------------------------------------------------------- f_{Lj3h}
Substituting Equations~(\ref{eq:ALj3h-final}) and (\ref{eq:MLj3h-pfin}) into Equation~(\ref{eq:fLj3h}) and ignoring correlations between $h_1$, $h_2$ and $h_3$ hadrons, we have
{\setlength\arraycolsep{0.2pt}
\begin{eqnarray}
 f_{L_{j}}(\bm{p})  &=& \frac{64 \pi^3 \hbar^6 c^6 g_{L_{j}} \gamma^2}{3\sqrt{3}(C_1 R_f^2+\sigma_1^2) \sqrt{C_1 (R_f/\gamma)^2+\sigma_1^2}}     \nonumber   \\
 && \times   \frac{1}{ (C_2 R_f^2+\sigma_2^2) \sqrt{C_2 (R_f/\gamma)^2+\sigma_2^2}}   f_{h_1}(\frac{m_1\bm{p}}{m_1+m_2+m_3})       \nonumber   \\
 && \times   f_{h_2}(\frac{m_2\bm{p}}{m_1+m_2+m_3})    f_{h_3}(\frac{m_3\bm{p}}{m_1+m_2+m_3}) .     \label{eq:fLj3h-approx}
\end{eqnarray} }%
Finally we have the Lorentz invariant momentum distribution
{\setlength\arraycolsep{0.2pt}
\begin{eqnarray}
 && f_{L_j}^{(inv)}(p_{T},y) = \frac{64 \pi^3 \hbar^6 c^6 g_{L_{j}} }{3\sqrt{3}(C_1 R_f^2+\sigma_1^2) \sqrt{C_1 (R_f/\gamma)^2+\sigma_1^2}}     \nonumber   \\
 && ~~~~~~ \times  \frac{1}{ (C_2 R_f^2+\sigma_2^2) \sqrt{C_2 (R_f/\gamma)^2+\sigma_2^2}}   \frac{m_1+m_2+m_3}{m_1m_2m_3}   \nonumber   \\
  &&  ~~~~~~ \times        f_{h_1}^{(inv)}(\frac{m_1p_{T}}{m_1+m_2+m_3},y)  f_{h_2}^{(inv)}(\frac{m_2p_{T}}{m_1+m_2+m_3},y)      \nonumber   \\
  &&  ~~~~~~ \times      f_{h_3}^{(inv)}(\frac{m_3p_{T}}{m_1+m_2+m_3},y) .     \label{eq:pt-Lj3h}
\end{eqnarray} }%

%--------------------------------------------------------------------------------------------------  ppnn --> He4   
\subsection{Formalism of four nucleons coalescing into $^4$He}

For $^4$He formed via the coalescence of four nucleons, the three-dimensional momentum distribution is
{\setlength\arraycolsep{0pt}
\begin{eqnarray}
&& f_{^4\text{He}}(\bm{p})= N_{ppnn} \int d\bm{x}_1d\bm{x}_2d\bm{x}_3d\bm{x}_4 d\bm{p}_1 d\bm{p}_2d\bm{p}_3d\bm{p}_4    \nonumber  \\
&&~~~~~~~~~~~~~~~~\times f^{(n)}_{ppnn}(\bm{x}_1,\bm{x}_2,\bm{x}_3,\bm{x}_4;\bm{p}_1,\bm{p}_2,\bm{p}_3,\bm{p}_4)  \nonumber  \\
&&~~~~~~~~~~~~~~~~\times \mathcal {R}_{^4\text{He}}(\bm{x}_1,\bm{x}_2,\bm{x}_3,\bm{x}_4;\bm{p}_1,\bm{p}_2,\bm{p}_3,\bm{p}_4,\bm{p}),     \label{eq:fHe4general1}               
\end{eqnarray} }%
where $N_{ppnn}=N_{p}(N_{p}-1)N_{n}(N_{n}-1)$ is the number of all possible $ppnn$-clusters;
$f^{(n)}_{ppnn}$ is the normalized four-nucleon joint coordinate-momentum distribution; $\mathcal {R}_{^4\text{He}}$ is the kernel function.

We rewrite the kernel function as
{\setlength\arraycolsep{0pt}
\begin{eqnarray}
&&  \mathcal {R}_{^4\text{He}}(\bm{x}_1,\bm{x}_2,\bm{x}_3,\bm{x}_4;\bm{p}_1,\bm{p}_2,\bm{p}_3,\bm{p}_4,\bm{p}) = g_{^4\text{He}}  \nonumber  \\
&& ~~~~   \times \mathcal {R}_{^4\text{He}}^{(x,p)}(\bm{x}_1,\bm{x}_2,\bm{x}_3,\bm{x}_4;\bm{p}_1,\bm{p}_2,\bm{p}_3,\bm{p}_4) \delta(\displaystyle{\sum^4_{i=1}} \bm{p}_i-\bm{p}),      \label{eq:RHe4fac}  
\end{eqnarray} }%
where the spin degeneracy factor $g_{^4\text{He}}=1/16$, and
{\setlength\arraycolsep{0pt}
\begin{eqnarray}
 && \mathcal {R}^{(x,p)}_{^4\text{He}}(\bm{x}_1,\bm{x}_2,\bm{x}_3,\bm{x}_4;\bm{p}_1,\bm{p}_2,\bm{p}_3,\bm{p}_4)=8^3e^{-\frac{(\bm{x}'_1-\bm{x}'_2)^2}{2\sigma_{^4\text{He}}^2}}   \nonumber  \\
&&~~~~~~~~~ \times  e^{-\frac{(\bm{x}'_1+\bm{x}'_2-2\bm{x}'_3)^2}{6\sigma_{^4\text{He}}^2}}  e^{-\frac{(\bm{x}'_1+\bm{x}'_2+\bm{x}'_3-3\bm{x}'_4)^2}{12\sigma_{^4\text{He}}^2}}   \nonumber  \\
&&~~~~~~~~~ \times e^{-\frac{\sigma_{^4\text{He}}^2(\bm{p}'_{1}-\bm{p}'_{2})^2}{2\hbar^2c^2}} e^{-\frac{\sigma_{^4\text{He}}^2(\bm{p}'_{1}+\bm{p}'_{2}-2\bm{p}'_{3})^2}{6\hbar^2c^2}}  
e^{-\frac{\sigma_{^4\text{He}}^2(\bm{p}'_{1}+\bm{p}'_{2}+\bm{p}'_{3}-3\bm{p}'_{4})^2}{12\hbar^2c^2}}.
\end{eqnarray} }%
Here $\sigma_{^4\text{He}}=\frac{2\sqrt{2}}{3}R_{^4\text{He}}$, and $R_{^4\text{He}}=1.6755$ fm~\cite{Angeli:2013epw} is the root-mean-square radius of the $^4$He.

Assuming that the normalized joint distribution is coordinate and momentum factorized,
we have
{\setlength\arraycolsep{0.2pt}
\begin{eqnarray}
 f_{^4\text{He}}(\bm{p}) 
= N_{ppnn} g_{^4\text{He}} \mathcal {A}_{^4\text{He}}   \mathcal {M}_{^4\text{He}}(\bm{p}).     \label{eq:fHe4}  
\end{eqnarray} }%
Here we use $\mathcal {A}_{^4\text{He}}$ to denote the coordinate integral part in Equation~(\ref{eq:fHe4}) as
{\setlength\arraycolsep{0pt}
\begin{eqnarray}
\mathcal {A}_{^4\text{He}} = && 8^3  \int d\bm{x}_1d\bm{x}_2d\bm{x}_3d\bm{x}_4 f^{(n)}_{ppnn}(\bm{x}_1,\bm{x}_2,\bm{x}_3,\bm{x}_4)    \nonumber  \\
&&~~~~~~ \times  e^{-\frac{(\bm{x}'_1-\bm{x}'_2)^2}{2\sigma_{^4\text{He}}^2}}  
         e^{-\frac{(\bm{x}'_1+\bm{x}'_2-2\bm{x}'_3)^2}{6\sigma_{^4\text{He}}^2}}  e^{-\frac{(\bm{x}'_1+\bm{x}'_2+\bm{x}'_3-3\bm{x}'_4)^2}{12\sigma_{^4\text{He}}^2}} ,      \label{eq:AHe4}      
\end{eqnarray} }%
and use $\mathcal {M}_{^4\text{He}}(\bm{p})$ to denote the momentum integral part as
{\setlength\arraycolsep{0pt}
\begin{eqnarray}
 &&  \mathcal {M}_{^4\text{He}}(\bm{p}) =   \int d\bm{p}_1d\bm{p}_2d\bm{p}_3d\bm{p}_4 f^{(n)}_{ppnn}(\bm{p}_1,\bm{p}_2,\bm{p}_3,\bm{p}_4) 
   \nonumber   \\
&& \times   e^{-\frac{\sigma_{^4\text{He}}^2(\bm{p}'_{1}-\bm{p}'_{2})^2}{2\hbar^2c^2}}
  e^{-\frac{\sigma_{^4\text{He}}^2(\bm{p}'_{1}+\bm{p}'_{2}-2\bm{p}'_{3})^2}{6\hbar^2c^2}}  
e^{-\frac{\sigma_{^4\text{He}}^2(\bm{p}'_{1}+\bm{p}'_{2}+\bm{p}'_{3}-3\bm{p}'_{4})^2}{12\hbar^2c^2}}
 \delta(\displaystyle{\sum^4_{i=1}} \bm{p}_i-\bm{p}).  \nonumber \\   \label{eq:MHe4}
\end{eqnarray} }%

%--------------------------------------------------------------------------------------------------- AHe4

We change integral variables in Equation~(\ref{eq:AHe4}) to be $\bm{Z}= (\bm{x}_1+\bm{x}_2+\bm{x}_3+\bm{x}_4)/2$, $\bm{r}_1= (\bm{x}_1-\bm{x}_2)/\sqrt{2}$, $\bm{r}_2= (\bm{x}_1+\bm{x}_2-2\bm{x}_3)/\sqrt{6}$ and $\bm{r}_3= (\bm{x}_1+\bm{x}_2+\bm{x}_3-3\bm{x}_4)/\sqrt{12}$, 
and assume $f^{(n)}_{ppnn}(\bm{Z},\bm{r}_1,\bm{r}_2,\bm{r}_3) = f^{(n)}_{ppnn}(\bm{Z}) f^{(n)}_{ppnn}(\bm{r}_1) f^{(n)}_{ppnn}(\bm{r}_2)$ $ f^{(n)}_{ppnn}(\bm{r}_3)$.
Adopting $f^{(n)}_{ppnn}(\bm{r}_1) = \frac{1}{(\pi C_1 R_f^2)^{3/2}} e^{-\frac{\bm{r}_1^2}{C_1 R_f^2}}$,
$f^{(n)}_{ppnn}(\bm{r}_2) = \frac{1}{(\pi C_2 R_f^2)^{3/2}} e^{-\frac{\bm{r}_2^2}{C_2 R_f^2}}$ and $f^{(n)}_{ppnn}(\bm{r}_3) = \frac{1}{(\pi C_3 R_f^2)^{3/2}} e^{-\frac{\bm{r}_3^2}{C_3 R_f^2}}$, we have
{\setlength\arraycolsep{0pt}
\begin{eqnarray}
 \mathcal {A}_{^4\text{He}} &=&  8^3\int d\bm{r}_1d\bm{r}_2d\bm{r}_3 f^{(n)}_{ppnn}(\bm{r}_1)f^{(n)}_{ppnn}(\bm{r}_2)f^{(n)}_{ppnn}(\bm{r}_3)      \nonumber  \\
    &&~~~~~~ \times   e^{-\frac{(\bm{r}'_1)^2}{\sigma_{^4\text{He}}^2}} e^{-\frac{(\bm{r}'_2)^2}{\sigma_{^4\text{He}}^2}} e^{-\frac{(\bm{r}'_3)^2}{\sigma_{^4\text{He}}^2}}.      \label{eq:AHe4-r}  
\end{eqnarray} }%
$C_1$, $C_2$, $C_3$ are equal to be $C_w$, $4C_w/3$ and $3C_w/2$, respectively ~\cite{Mrowczynski:2016xqm,Wang:2020zaw}. 
After the Lorentz transformation and integrating the relative coordinate variables from Equation~(\ref{eq:AHe4-r}), we obtain 
{\setlength\arraycolsep{0pt}
\begin{eqnarray}
  \mathcal {A}_{^4\text{He}} &=&  \frac{8^3 \sigma_{^4\text{He}}^9} {(C_1 R_f^2+\sigma_{^4\text{He}}^2) \sqrt{C_1 (R_f/\gamma)^2+\sigma_{^4\text{He}}^2} }   \nonumber  \\
 && \times \frac{1}{(C_2 R_f^2+\sigma_{^4\text{He}}^2) \sqrt{C_2 (R_f/\gamma)^2+\sigma_{^4\text{He}}^2} }   \nonumber  \\
 && \times  \frac{1}{(C_3 R_f^2+\sigma_{^4\text{He}}^2) \sqrt{C_3 (R_f/\gamma)^2+\sigma_{^4\text{He}}^2}} .   \label{eq:AHe4-fin}  
\end{eqnarray} }%

%--------------------------------------------------------------------------------------------------- MHe4
Approximating the gaussian form of the momentum-dependent kernel function to be $\delta$ function form and after integrating $\bm{p}_1$, $\bm{p}_2$, $\bm{p}_3$ and $\bm{p}_4$ in Equation~(\ref{eq:MHe4}),  we can obtain
{\setlength\arraycolsep{0.2pt}
\begin{eqnarray}
 \mathcal {M}_{^4\text{He}}(\bm{p}) &=&   (\frac{\pi^{3/2}\hbar^3 c^3}{2\sigma_{^4\text{He}}^3})^3 \gamma^3  f^{(n)}_{p}(\frac{\bm{p}}{4}) f^{(n)}_{p}(\frac{\bm{p}}{4})
  f^{(n)}_{n}(\frac{\bm{p}}{4})   f^{(n)}_{n}(\frac{\bm{p}}{4})  .~~~~~~    \label{eq:MHe4-fin}  
\end{eqnarray} }%

%--------------------------------------------------------------------------------------------------- f_He4
Substituting Equations~(\ref{eq:AHe4-fin}) and (\ref{eq:MHe4-fin}) into Equation~(\ref{eq:fHe4}), we have
{\setlength\arraycolsep{0.2pt}
\begin{eqnarray}
 f_{^4\text{He}}(\bm{p}) &=&\frac{64  g_{^4\text{He}}  \gamma^3 \pi^{9/2}\hbar^9 c^9}
 {(C_1 R_f^2+\sigma_{^4\text{He}}^2) \sqrt{C_1 (R_f/\gamma)^2+\sigma_{^4\text{He}}^2} }    \nonumber  \\
&& \times   \frac{1}{(C_2 R_f^2+\sigma_{^4\text{He}}^2) \sqrt{C_2 (R_f/\gamma)^2+\sigma_{^4\text{He}}^2} }   \nonumber  \\
&& \times   \frac{1}{ (C_3 R_f^2+\sigma_{^4\text{He}}^2) \sqrt{C_3 (R_f/\gamma)^2+\sigma_{^4\text{He}}^2}}    \nonumber  \\
&& \times   f_{p}(\frac{\bm{p}}{4}) f_{p}(\frac{\bm{p}}{4}) f_{n}(\frac{\bm{p}}{4}) f_{n}(\frac{\bm{p}}{4}) .  \label{eq:fHe4-approx}
\end{eqnarray} }%
We finally have the Lorentz invariant momentum distribution
{\setlength\arraycolsep{0.2pt}
\begin{eqnarray}
 && f_{^4\text{He}}^{(inv)}(p_{T},y) = \frac{256  g_{^4\text{He}}  \pi^{9/2}\hbar^9 c^9}
 {m^3 (C_1 R_f^2+\sigma_{^4\text{He}}^2) \sqrt{C_1 (R_f/\gamma)^2+\sigma_{^4\text{He}}^2} }    \nonumber  \\
&&~~ \times   \frac{1}{(C_2 R_f^2+\sigma_{^4\text{He}}^2) \sqrt{C_2 (R_f/\gamma)^2+\sigma_{^4\text{He}}^2} }   \nonumber  \\
&&~~ \times   \frac{1}{ (C_3 R_f^2+\sigma_{^4\text{He}}^2) \sqrt{C_3 (R_f/\gamma)^2+\sigma_{^4\text{He}}^2}}      \nonumber   \\
  &&~~  \times      f_{p}^{(inv)}(\frac{p_{T}}{4},y) f_{p}^{(inv)}(\frac{p_{T}}{4},y) f_{n}^{(inv)}(\frac{p_{T}}{4},y) f_{n}^{(inv)}(\frac{p_{T}}{4},y) , ~~~    \label{eq:pt-He4}
\end{eqnarray} }%
where $m$ is the nucleon mass.

As a short summary of this section, we want to state that Equations~(\ref{eq:pt-Lj2h}, \ref{eq:pt-Lj3h}, \ref{eq:pt-He4}) show the relationship of light nuclei with primordial hadronic bodies in momentum space in the laboratory frame. 
They can be directly used to calculate the yields and $p_T$ distributions of light nuclei formed via different coalescence channels as long as the primordial hadronic momentum distributions are given.
In the case of ignoring the mass differences of primordial hadrons, Equations~(\ref{eq:pt-Lj2h}) and (\ref{eq:pt-Lj3h}) return to our previous results for $d$, $t$ and $^3$He in Refs.~\cite{Wang:2020zaw,Zhao:2022xkz} where only nucleon coalescence was considered.

%----------------------------------------------------------------------------------------------------------------------------------------------------------------------------------------------------------------- applications  
\section{Results and discussions} \label{results}

In this section, we apply the coalescence model in Sec.~\ref{model} to Au-Au collisions at $\sqrt{s_{NN}}=3$ GeV to study the momentum and centrality dependence of the production of different light nuclei in the low- and intermediate-$p_T$ regions in different rapidity intervals at midrapidity area.
We first introduce the $p_T$ spectra of the nucleons. 
We then present $p_T$ dependence of different coalescence sources for $d$, $t$, $^3$He and $^4$He in their productions.
We finally give the yield rapidity densities $dN/dy$ and the averaged transverse momenta $\langle p_T \rangle$ of different light nuclei.

%------------------------------------------------------------------------------------------------------------------------------ nucleon pt spectra
\subsection{$p_T$ spectra of nucleons}

The invariant $p_T$ distributions at different rapidity intervals of  primordial protons $ f_{p,\text{pri}}^{(\text{inv})}(p_{T},y) $ and neutrons $ f_{n,\text{pri}}^{(\text{inv})}(p_{T},y) $ are necessary inputs for computing $p_T$ distributions of light nuclei in our model.
The relationship of primordial protons and those final-state ones is as follows
{\setlength\arraycolsep{0pt}
\begin{eqnarray}
f_{p,\text{pri}}^{(\text{inv})}(p_{T},y) - f_{p,\text{lignucl}}^{(\text{inv})}(p_{T},y) + f_{p,\text{hypdec}}^{(\text{inv})}(p_{T},y)=&& f_{p,\text{fin}}^{(\text{inv})}(p_{T},y).    \nonumber   \\   \label{eq:fp}  
\end{eqnarray} }%
The last three terms in the equation denote the invariant $p_T$ distributions of protons consumed in light nuclei production, those coming from hyperon weak decays and those final-state ones, respectively.
The feed-down contribution from the weak decays of hyperons to protons is about $1.5\%$~\cite{Liu:2022ump} and that entering into light nuclei takes about $20\%$~\cite{Liu:2021xfy}.
Considering that most of primordial protons, more than 80\%, evolve to be final-state ones, we ignore the variation of the shape of the $p_T$ spectra of primordial protons and final-state ones.
In this case, we can get $f_{p,\text{pri}}^{(\text{inv})}(p_{T},y) \approx \frac{1}{81.5\%}f_{p,\text{fin}}^{(\text{inv})}(p_{T},y)$.

We here use the blast-wave model to get invariant $p_T$ distribution functions of final-state protons by fitting the proton experimental data in Ref.~\cite{Liu:2021xfy}.
The blast-wave function~\cite{Schnedermann:1993ws} is given as
{\setlength\arraycolsep{0.2pt}
\begin{eqnarray}
f_{p,fin}^{(inv)}(p_{T},y) = \frac{d^{2}N_{p,fin}}{2\pi p_{T}dp_{T}dy}  \propto  &&  \int_{0}^{R} r dr m_T I_0 \left(\frac{p_Tsinh\rho}{T_{kin}}\right)     \nonumber   \\
    && ~~~~~ \times  K_1\left(\frac{m_Tcosh\rho}{T_{kin}}\right) ,  \label{eq:BWfitfunc}
\end{eqnarray} }%
where $r$ is the radial distance in the transverse plane and $R$ is the radius of the fireball.
$m_T$ is the transverse mass of the proton. $I_0$ and $K_1$ are the modified Bessel functions, and the velocity profile $\rho=tanh^{-1}[\beta_s(\frac{r}{R})^n]$.
The surface velocity $\beta_s$, the kinetic freeze-out temperature $T_{kin}$ and $n$ are fitting parameters.

\begin{figure*}[htbp]
\centering
 \includegraphics[width=0.7\linewidth]{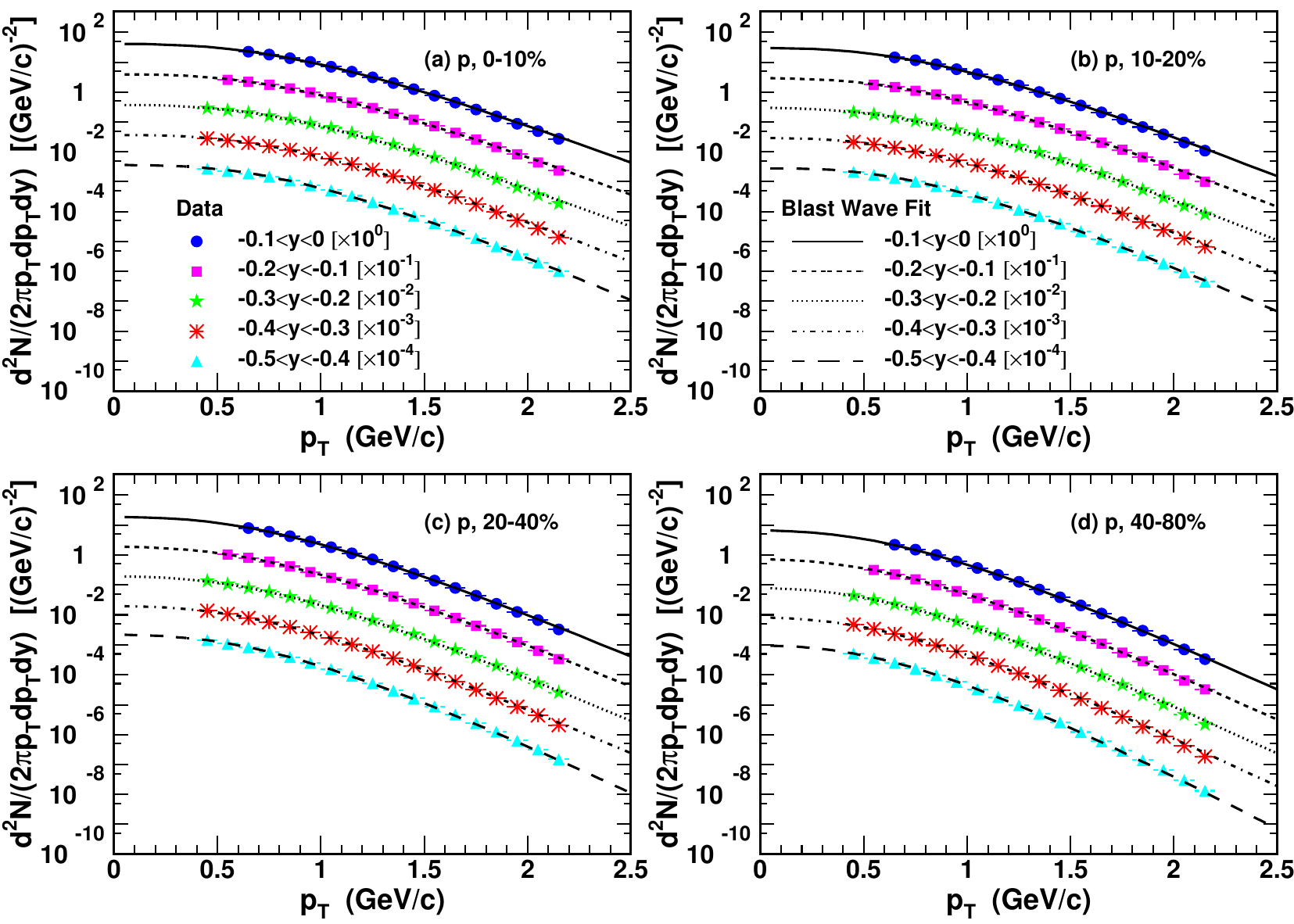}\\
 \caption{(Color online) Invariant $p_T$ spectra of protons in different rapidity intervals at midrapidity in Au-Au collisions at $\sqrt{s_{NN}}=3$ GeV in centralities (a) $0-10$\%, (b) $10-20$\%, (c) $20-40$\%, (d) $40-80$\%. Filled symbols are experimental data~\cite{Liu:2021xfy}. Different lines are the fitting results by the blast-wave model.
 }
 \label{fig:ppT}
\end{figure*}

Figure \ref{fig:ppT} shows the invariant $p_T$ spectra of final-state protons in different rapidity intervals $-0.1<y<0$, $-0.2<y<-0.1$, $-0.3<y<-0.2$, $-0.4<y<-0.3$, $-0.5<y<-0.4$ in Au-Au collisions at $\sqrt{s_{NN}}=3$ GeV in centralities $0-10$\%, $10-20$\%, $20-40$\%, $40-80$\%.
The spectra in different rapidity intervals are scaled by different factors for clarity as shown in the figure.
Filled symbols are experimental data from the STAR collaboration~\cite{Liu:2021xfy}. 
Different lines are the results of the blast-wave model.
Since we focus on testing the validity of the coalescence mechanism in describing the light nuclei production at low collision energy instead of predicting the momentum distributions of light nuclei,
we only include the best fit from the blast-wave model for the proton, and do not consider the fitting errors.
Here, the proton $dN/dy$ and $\langle p_T \rangle$ obtained by these blast-wave results are just equal to central values of the corresponding data given by the STAR collaboration in Ref.~\cite{Liu:2021xfy}.  

\begin{table}[htbp]
	\centering
	\caption{Values of $Z_{np}$ and $R_f$ in different rapidity intervals and different centralities in Au-Au collisions at $\sqrt{s_{NN}}=3$ GeV.}
	\begin{tabular}{cccc}
		\toprule
		Centrality                               &Rapidity                   &$~~~~~Z_{np}~~~~~$        &$R_f$ (fm)  \\
		\hline
		\multirow{5}{*}{$0-10\%$}   &$-0.1<y<0~~~~~$    &$1.33$             &$3.30$    \\
		                                              &$-0.2<y<-0.1$          &$1.33$             &$3.21$  \\
		                                              &$-0.3<y<-0.2$          &$1.29$             &$3.10$  \\
		                                              &$-0.4<y<-0.3$          &$1.34$             &$3.09$  \\
		                                              &$-0.5<y<-0.4$          &$1.36$             &$3.08$  \\
		 \hline
			
		\multirow{5}{*}{$10-20\%$}  &$-0.1<y<0~~~~~$    &$1.34$             &$2.93$   \\
		                                              &$-0.2<y<-0.1$          &$1.34$             &$2.84$  \\
		                                              &$-0.3<y<-0.2$          &$1.27$             &$2.78$  \\
		                                              &$-0.4<y<-0.3$          &$1.30$             &$2.76$  \\
		                                              &$-0.5<y<-0.4$          &$1.33$             &$2.70$  \\
		 \hline
		
		\multirow{5}{*}{$20-40\%$} &$-0.1<y<0~~~~~$    &$1.35$             &$2.51$    \\
		                                              &$-0.2<y<-0.1$          &$1.32$             &$2.42$  \\
		                                              &$-0.3<y<-0.2$          &$1.25$             &$2.35$  \\
		                                              &$-0.4<y<-0.3$          &$1.28$             &$2.34$  \\
		                                              &$-0.5<y<-0.4$          &$1.28$             &$2.32$  \\
		 \hline
		
		\multirow{5}{*}{$40-80\%$} &$-0.1<y<0~~~~~$    &$1.15$             &$1.42$    \\
		                                              &$-0.2<y<-0.1$          &$1.13$             &$1.41$  \\
		                                              &$-0.3<y<-0.2$          &$1.01$             &$1.40$  \\
		                                              &$-0.4<y<-0.3$          &$1.05$            &$1.35$  \\
		                                              &$-0.5<y<-0.4$          &$1.05$            &$1.34$  \\

		\hline\hline
		\label{tab:AnpRf}
	\end{tabular}
\end{table}

For the neutron, we assume the same normalized $p_T$ distribution as that of the proton in the same rapidity interval and the same collision centrality.
For absolute yield density of the neutron, it is generally not equal to that of the proton due to the prominent influences of net nucleons from the colliding Au nuclei.
We here use $Z_{np}$ to denote the extent of the yield density asymmetry of the neutron and the proton and take their relation as
{\setlength\arraycolsep{0pt}
\begin{eqnarray}
 \frac{dN_n}{dy}=\frac{dN_p}{dy}\times Z_{np}.   \label{eq:NnNprelation}  
\end{eqnarray} }%
$Z_{np}=1$ corresponds to the complete isospin equilibration and $Z_{np}=1.49$ to isospin asymmetry in the whole Au nucleus.
We here set $Z_{np}$ to be a free parameter, and its values in different centrality and rapidity windows are put in Table \ref{tab:AnpRf}.
Values of $Z_{np}$ in central and semi-central $0-10$\%, $10-20$\%, $20-40$\%  centralities are comparable and they are close to that evaluated in Ref.~\cite{Kittiratpattana:2022knq}.
$Z_{np}$ in $40-80$\% centrality becomes a little smaller.
From the viewpoint of the effect of the neutron skin~\cite{V:2022zqk}, $Z_{np}$ is expected to increase in peripheral collisions.
But note that we here study light nuclei production in the midrapidity area, i.e., $y<0.5$, in peripheral collisions the transparency of nucleons from the colliding nuclei become stronger due to smaller reaction area and they move to relative larger rapidity~\cite{Liu:2021xfy}.
The participant nucleons from colliding nuclei become less in midrapidity region, so the yield asymmetry extent due to the participant nucleons decreases.

The other parameter in our model is $R_f$, which is fixed by the data of the yield rapidity density of $d$~\cite{Liu:2021xfy}.
Values of $R_f$ in different rapidity intervals and different centralities in Au-Au collisions at $\sqrt{s_{NN}}=3$ GeV are listed in Table \ref{tab:AnpRf}.
For 0-10\% centrality, our fixed values locate in the range evaluated by the linear dependence on the cube root of the rapidity density of charged particles, i.e., $R_f \propto (dN_{ch}/dy)^{1/3}$ ~\cite{Zhao:2022xkz,Kimelman:2022hwp}.
For other collision centralities, $R_f$ cannot be evaluated by the relation $R_f \propto (dN_{ch}/dy)^{1/3}$ due to the lack of the data of $\pi^{\pm}$ and $K^{\pm}$ currently.
From Table \ref{tab:AnpRf}, one can see $R_f$ decreases very slightly as the increasing rapidity for the same centrality and it decreases from central to peripheral collisions.
The smaller $R_f$ in more peripheral collisions leads to the stronger suppression of light nuclei production because of the non-negligible light nuclei sizes compared to $R_f$ as shown in Equations~(\ref{eq:pt-Lj2h}, \ref{eq:pt-Lj3h}, \ref{eq:pt-He4}).
This suppression effect of light nuclei production in small collision systems has been systematically studied in Ref.~\cite{Sun:2018mqq}.

%------------------------------------------------------------------------------------------------------------------------------ light nuclei pt spectra
\subsection{$p_T$ spectra of light nuclei}

\begin{figure*}[htbp]
\centering
 \includegraphics[width=0.7\linewidth]{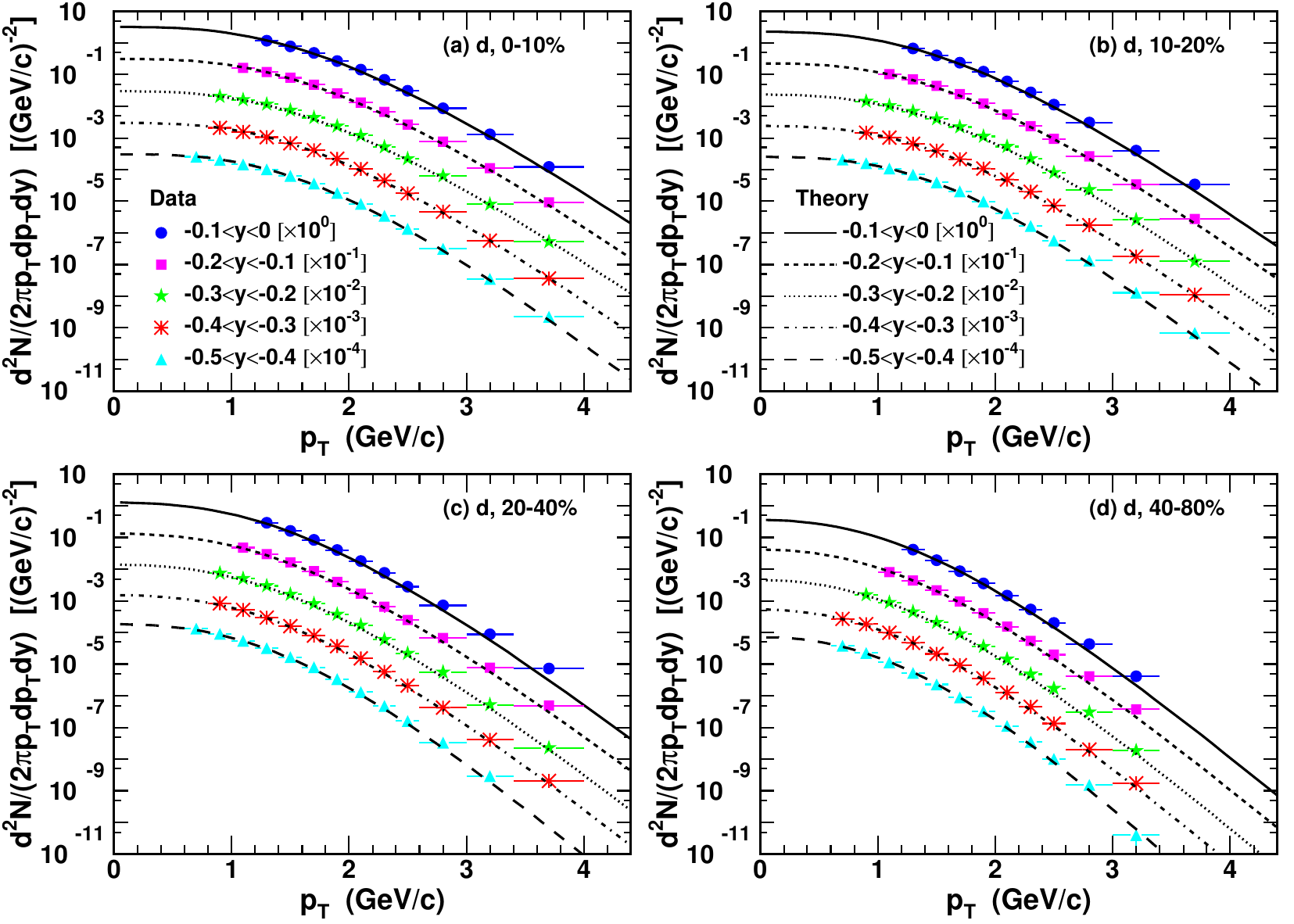}\\
 \caption{(Color online) Invariant $p_T$ spectra of deuterons in different rapidity intervals at midrapidity in Au-Au collisions at $\sqrt{s_{NN}}=3$ GeV in centralities (a) $0-10$\%, (b) $10-20$\%, (c) $20-40$\%, (d) $40-80$\%. Filled symbols are experimental data~\cite{Liu:2021xfy}. Different lines are the theoretical results for final-state deuterons.
 }
 \label{fig:dpT}
\end{figure*}

With Equation (\ref{eq:pt-Lj2h}), we first compute the invariant $p_T$ distributions of deuterons in rapidity intervals $-0.1<y<0$, $-0.2<y<-0.1$, $-0.3<y<-0.2$, $-0.4<y<-0.3$, $-0.5<y<-0.4$ in Au-Au collisions at $\sqrt{s_{NN}}=3$ GeV in centralities $0-10$\%, $10-20$\%, $20-40$\%, $40-80$\%, respectively.
Here, $h_1$ and $h_2$ in Equation (\ref{eq:pt-Lj2h}) refer to the proton and the neutron.
Different lines scaled by different factors for clarity in Figure \ref{fig:dpT} are our theoretical results for final-state deuterons, i.e., those obtained by subtracting consumed ones in the nucleus coalescence from formed ones via the $p+n$ coalescence.
Filled symbols with error bars are experimental data from the STAR collaboration~\cite{Liu:2021xfy}. 
From Figure \ref{fig:dpT}, one can see our results can well reproduce the available data in different rapidity intervals at midrapidity area from central to peripheral Au-Au collisions at $\sqrt{s_{NN}}=3$ GeV.

\begin{figure*}[htbp]
\centering
 \includegraphics[width=0.8\linewidth]{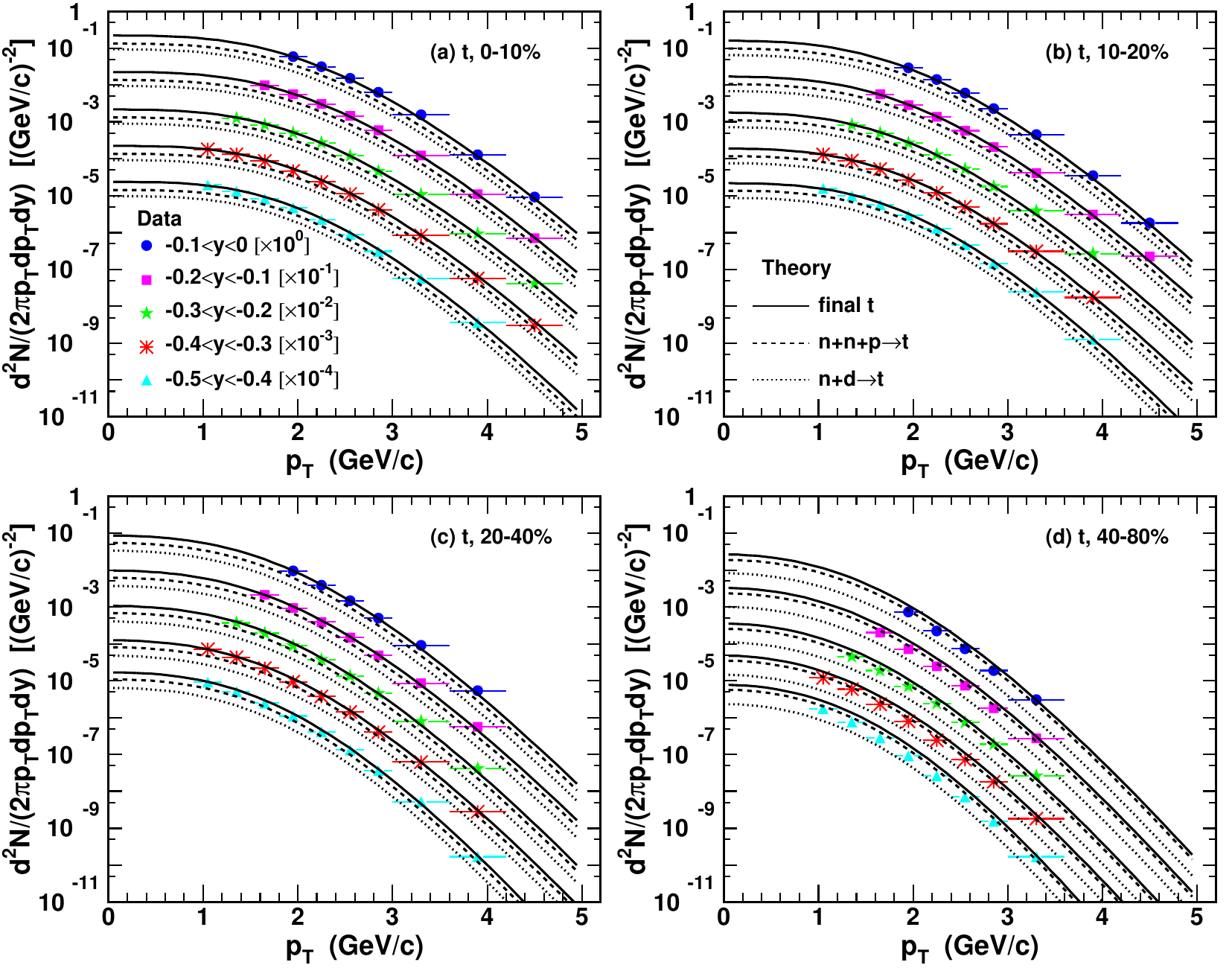}\\
 \caption{(Color online) Invariant $p_T$ spectra of tritons in different rapidity intervals at midrapidity in Au-Au collisions at $\sqrt{s_{NN}}=3$ GeV in centralities (a) $0-10$\%, (b) $10-20$\%, (c) $20-40$\%, (d) $40-80$\%. Filled symbols are experimental data~\cite{Liu:2021xfy}. Solid, dashed and dotted lines are the theoretical results of final tritons, $n+n+p$ coalescence and $n+d$ coalescence, respectively.
 }
 \label{fig:tpT}
\end{figure*}

We then study the invariant $p_T$ distributions of $t$, $^3$He and $^4$He in rapidity intervals $-0.1<y<0$, $-0.2<y<-0.1$, $-0.3<y<-0.2$, $-0.4<y<-0.3$, $-0.5<y<-0.4$ in Au-Au collisions at $\sqrt{s_{NN}}=3$ GeV in centralities $0-10$\%, $10-20$\%, $20-40$\%, $40-80$\%, respectively.
Figure \ref{fig:tpT} shows the invariant $p_T$ spectra of tritons.
The spectra in different rapidity intervals are scaled by different factors for clarity as shown in the figure.
Filled symbols with error bars are experimental data from the STAR collaboration~\cite{Liu:2021xfy}. 
Dashed lines are the results of the nucleon coalescence, i.e., the contribution of the channel $n+n+p \rightarrow t$. 
Dotted lines are the results of the $n+d$ coalescence.
Solid lines are the final results of the $n+n+p $ coalescence plus $n+d$ coalescence minus consumed in $p+t$ coalescence.
Panels (a), (b) and (c) in Figure \ref{fig:tpT} show results of $n+n+p$ coalescence plus $n+d$ coalescence minus consumed in $p+t$ coalescence can describe the available data well in central and semi-central Au-Au collisions at $\sqrt{s_{NN}}=3$ GeV,
while panel (d) shows triton production in peripheral $40-80$\% Au-Au collisions favors $n+n+p$ coalescence. 

\begin{figure*}[htbp]
\centering
 \includegraphics[width=0.85\linewidth]{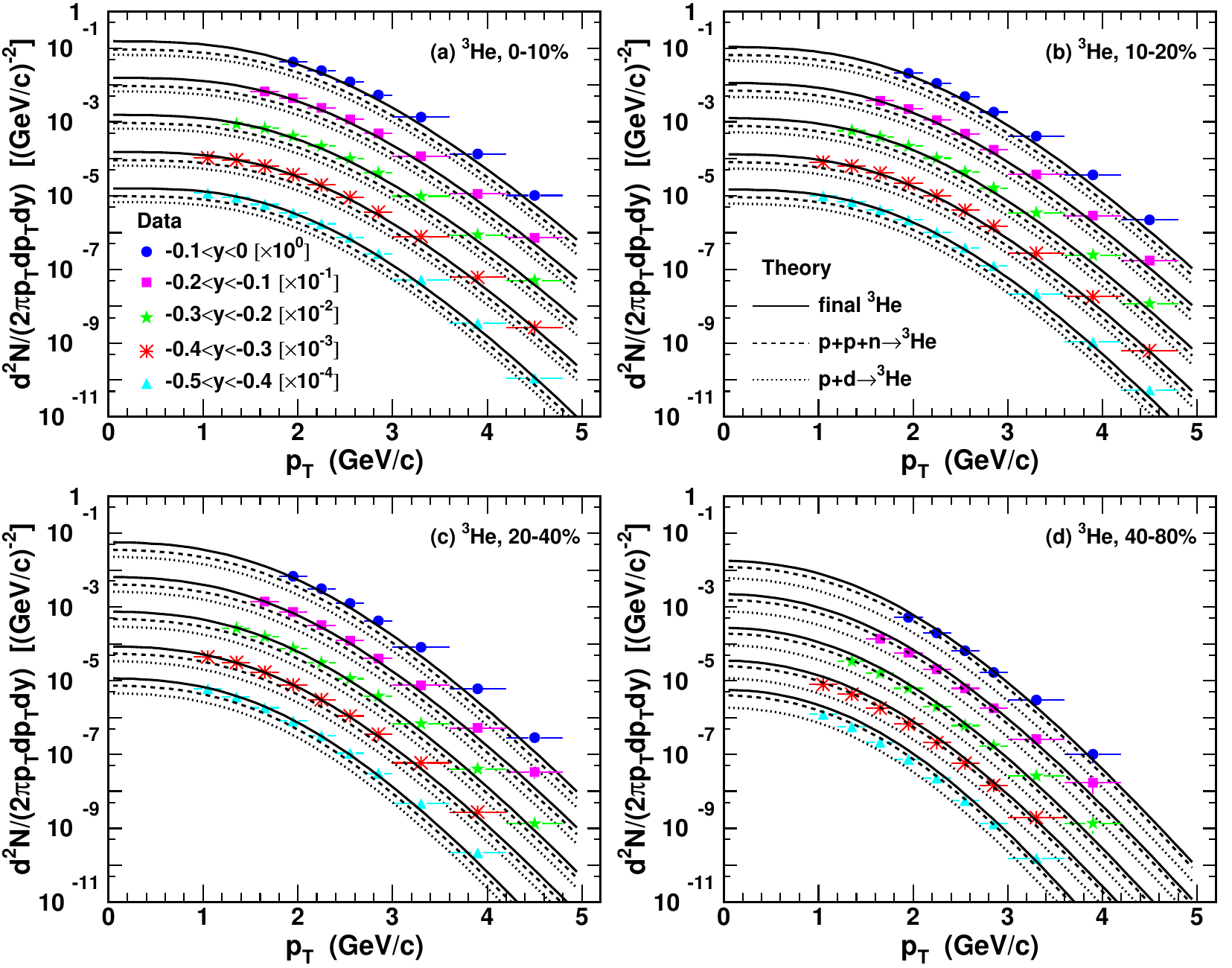}\\
 \caption{(Color online) Invariant $p_T$ spectra of $^3$He in different rapidity intervals at midrapidity in Au-Au collisions at $\sqrt{s_{NN}}=3$ GeV in centralities (a) $0-10$\%, (b) $10-20$\%, (c) $20-40$\%, (d) $40-80$\%. Filled symbols are experimental data~\cite{Liu:2021xfy}. Solid, dashed and dotted lines are the theoretical results of final $^3$He's, $p+p+n $ coalescence and $p+d$ coalescence, respectively.
 }
 \label{fig:He3pT}
\end{figure*}

Figure \ref{fig:He3pT} shows the invariant $p_T$ spectra of $^3$He.
The spectra in different rapidity intervals are also scaled by different factors for clarity as shown in the figure.
Filled symbols with error bars are experimental data from the STAR collaboration~\cite{Liu:2021xfy}. 
Dashed lines are the results of the nucleon coalescence, i.e., the contribution of the channel $p+p+n \rightarrow ^3$He. 
Dotted lines are the results of the $p+d$ coalescence.
Solid lines are the final results of the $p+p+n $ coalescence plus $p+d$ coalescence minus consumed in $n+^3$He coalescence.
From panels (a), (b) and (c) in Figure \ref{fig:He3pT}, one can see results of $p+p+n $ coalescence plus $p+d$ coalescence minus consumed in $n+^3$He coalescence can describe the available data well in central and semi-central Au-Au collisions at $\sqrt{s_{NN}}=3$ GeV.
But panel (d) in Figure \ref{fig:He3pT} shows that $^3$He production in peripheral Au-Au collisions favors $p+p+n $ coalescence. 
This is similar to that of the triton.

\begin{figure*}[htbp]
\centering
 \includegraphics[width=0.88\linewidth]{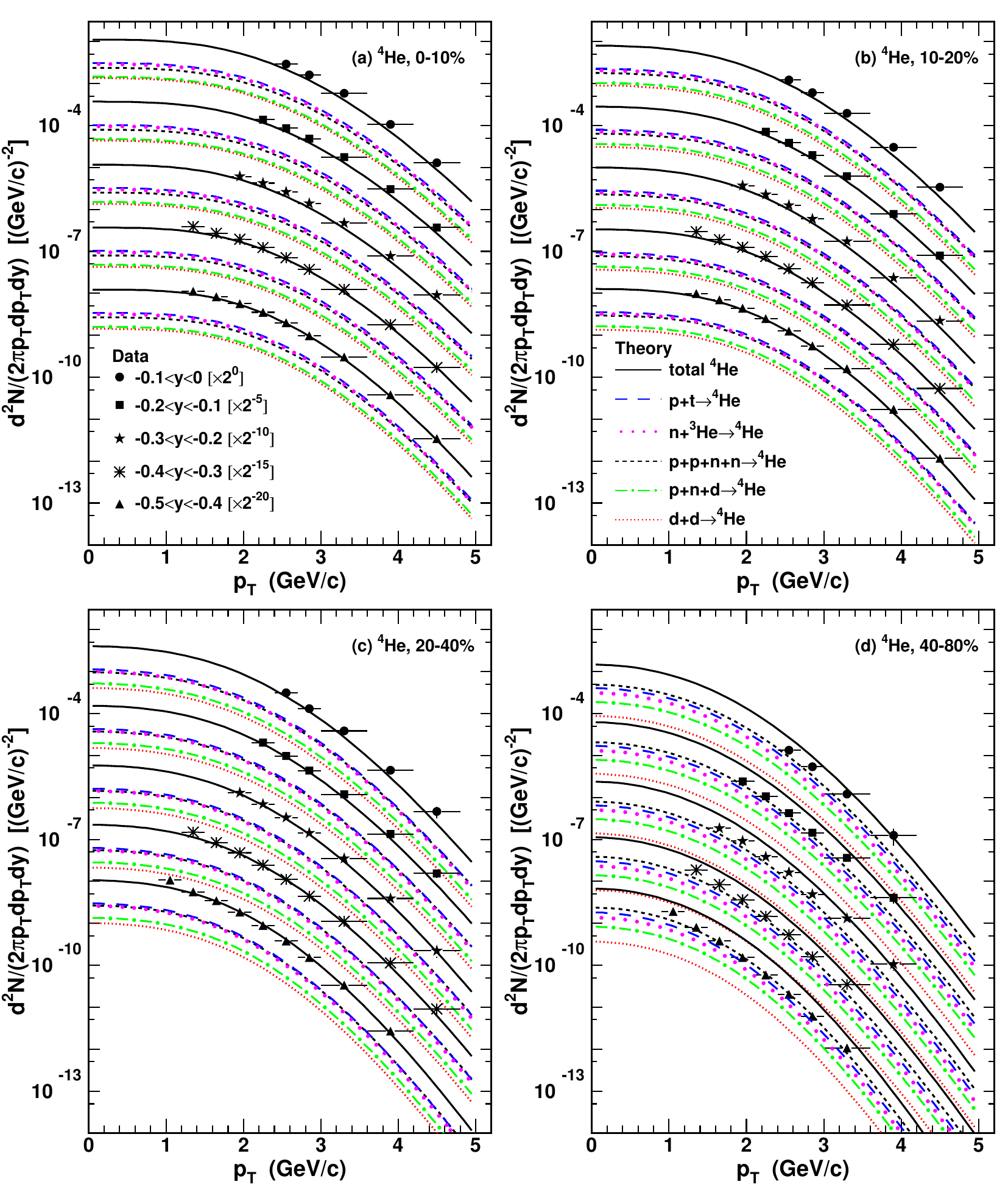}\\
 \caption{(Color online) Invariant $p_T$ spectra of $^4$He in different rapidity intervals at midrapidity in Au-Au collisions at $\sqrt{s_{NN}}=3$ GeV in centralities (a) $0-10$\%, (b) $10-20$\%, (c) $20-40$\%, (d) $40-80$\%. Filled symbols are experimental data~\cite{Liu:2021xfy}. Different lines are the theoretical results.
 }
 \label{fig:He4pT}
\end{figure*}

Figure \ref{fig:He4pT} shows the invariant $p_T$ spectra of $^4$He.
The spectra in different rapidity intervals are scaled by different factors for clarity as shown in the figure.
Filled symbols with error bars are experimental data from the STAR collaboration~\cite{Liu:2021xfy}. 
Short-dashed lines are the results of the nucleon coalescence, i.e., the contribution of the channel $p+p+n+n \rightarrow ^4$He. 
Long-dashed lines are the results of the contributions from the channel $p+t \rightarrow ^4$He, and large-gap dotted lines are the results of the contributions from the channel $n+^3$He$\rightarrow ^4$He.
Small-gap dotted lines are the results of the contributions from the channel $d+d\rightarrow ^4$He.
Dashed-dotted lines are the results of the contributions from the channel $p+n+d\rightarrow ^4$He.
Solid lines are the total results including the above five coalescence channels.
From panels (a), (b) and (c) in Figure \ref{fig:He4pT}, one can see total results including the above five coalescence processes can describe the available data in central and semi-central Au-Au collisions at $\sqrt{s_{NN}}=3$ GeV.
But panel (d) in Figure \ref{fig:He4pT} shows that $^4$He production in peripheral Au-Au collisions favors nucleon coalescence, i.e., $p+p+n+n$ coalescence. 
The other four coalescence cases involving nucleon$+$nucleus or nucleus$+$nucleus coalescence may not occur.

When calculating contributions from different coalescence channels, we base on the hypothesis that the nucleon coalescence happens first and subsequently the formed lighter cluster captures other particle to form heavier cluster if they meet the coalescence requirements in the phase space.
This coalescence time order is constrained to the local freeze-out instead of the whole phase space.
Results in Figures  \ref{fig:dpT}, \ref{fig:tpT} and \ref{fig:He3pT} show that 
our final results of $p_T$ spectra of $d$, $t$ and $^3$He can describe the experimental data in $0-10$\%, $10-20$\%, $20-40$\% centralities while in $40-80$\% centrality our results of nucleon coalescence itself can reproduce the available data.
Results in Figure \ref{fig:He4pT} show that our total results of nucleon coalescence plus nucleon$+d$ ($t$, $^3$He) coalescence plus $d+d$ coalescence can describe the data of $p_T$ spectra of $^4$He in $0-10$\%, $10-20$\%, $20-40$\% centralities while in $40-80$\% centrality nucleon coalescence itself can reproduce the $^4$He data.
This indicates that besides nucleon coalescence, nucleon/nucleus$+$nucleus coalescence plays an important role in central and semicentral collisions. 
But in peripheral collisions nucleus coalescence seems to disappear.
This is probably due to that the interactions between hadronic rescatterings become not so strong that the formed light nuclei can not capture other particles to form heavier objects.

%------------------------------------------------------------------------------------------------------------------------------ light nuclei dN/dy
\subsection{Yield rapidity densities of light nuclei}

\begin{table}[htbp]
	\centering
	\caption{Yield rapidity densities $dN/dy$ of $d$ in different rapidity intervals and different centralities in Au-Au collisions at $\sqrt{s_{NN}}=3$ GeV.
Data in the third column are from Ref.~\cite{Liu:2021xfy}, and the errors denote the systematical uncertainties.
Theo$_{pn}$, Theo$_{\text{dep}}$ and Theo$_{\text{fin}}$ in the last three columns denote those from $pn$ coalescence, depletions in forming $t$, $^3$He, $^4$He, and final-state ones, respectively.}
	\begin{tabular}{cccccc}
		\toprule
	\multirow{2}{*}{Centrality}  &\multirow{2}{*}{Rapidity}  &                      &                        &$d$                                       &                               \\
                                                                                                 \cline{3-6}
		                                      &                                           &Data                &Theo$_{pn}$   &Theo$_{\text{dep}}$   &Theo$_{\text{fin}}$          \\
      \hline
	\multirow{5}{*}{$0-10\%$}  &$-0.1<y<0~~~~~$             &$16.21\pm1.45$ &17.80               &1.45                               &16.35                            \\
		                                      &$-0.2<y<-0.1$                   &$16.07\pm1.42$ &17.45               &1.47                               &15.98                            \\
		                                      &$-0.3<y<-0.2$                   &$15.26\pm1.12$ &16.02               &1.37                               &14.65                            \\
		                                      &$-0.4<y<-0.3$                   &$14.69\pm1.19$ &16.04               &1.40                               &14.64                            \\
		                                      &$-0.5<y<-0.4$                   &$14.17\pm0.96$ &15.72               &1.40                               &14.32                            \\
	\hline

    \multirow{5}{*}{$10-20\%$}  &$-0.1<y<0~~~~~$             &$9.36\pm1.02$   &10.46              &0.82                               &9.64         \\
		                                      &$-0.2<y<-0.1$                   &$9.64\pm1.11$   &10.50              &0.86                               &9.64         \\
		                                      &$-0.3<y<-0.2$                   &$9.65\pm0.72$   &10.29              &0.87                               &9.42         \\
		                                      &$-0.4<y<-0.3$                   &$9.83\pm0.75$   &10.54              &0.92                               &9.62         \\
		                                      &$-0.5<y<-0.4$                   &$10.54\pm0.85$ &11.05              &1.04                               &10.01         \\
	\hline
			
    \multirow{5}{*}{$20-40\%$}  &$-0.1<y<0~~~~~$             &$4.59\pm0.53$  &4.79               &0.33                               &4.46       \\
		                                      &$-0.2<y<-0.1$                   &$4.84\pm0.65$  &4.92               &0.36                               &4.56       \\
		                                      &$-0.3<y<-0.2$                   &$5.05\pm0.50$  &5.00               &0.39                               &4.61       \\
		                                      &$-0.4<y<-0.3$                   &$5.30\pm0.50$  &5.39               &0.44                               &4.95       \\
		                                      &$-0.5<y<-0.4$                   &$6.09\pm0.42$  &6.16               &0.57                               &5.59       \\
	\hline

    \multirow{5}{*}{$40-80\%$}  &$-0.1<y<0~~~~~$             &$0.88\pm0.12$  &0.90               &0.05                               &0.85       \\
		                                      &$-0.2<y<-0.1$                   &$0.99\pm0.05$  &1.00               &0.07                               &0.93       \\
		                                      &$-0.3<y<-0.2$                   &$1.07\pm0.10$  &1.03               &0.07                               &0.96       \\
		                                      &$-0.4<y<-0.3$                   &$1.25\pm0.07$  &1.17               &0.09                               &1.08       \\
		                                      &$-0.5<y<-0.4$                   &$1.60\pm0.11$  &1.53               &0.14                               &1.39       \\

		\hline\hline
		\label{tab:dNdy-d}
	\end{tabular}
\end{table}

\begin{table*}[htbp]
	\centering
	\caption{Yield rapidity densities $dN/dy$ of $t$ and $^3$He in different rapidity intervals and different centralities in Au-Au collisions at $\sqrt{s_{NN}}=3$ GeV.
Data are from Ref.~\cite{Liu:2021xfy}, and the errors denote the systematical uncertainties.}
	\begin{tabular}{cccccccccccccc}
		\toprule
	\multirow{2}{*}{Centrality}  &\multirow{2}{*}{Rapidity}  &       &                 &$t$     &     &     &                                                                                 &         &                    &$^3$He                   &    &                 \\
                                                                                                 \cline{3-7}                                                                                                                              \cline{9-13}
		                                      &                                         &Data &Theo$_{nnp}$ &Theo$_{nd}$  &Theo$_{\text{dep}}$ &Theo$_{\text{fin}}$ &     &Data &Theo$_{ppn}$ &Theo$_{pd}$ &Theo$_{\text{dep}}$ &Theo$_{\text{fin}}$ &             \\
      \hline
	\multirow{5}{*}{$0-10\%$}  &$-0.1<y<0~~~~~$             &$1.968\pm0.218$ &1.199 &0.820 &0.042 &1.977   &                                                           &$1.356\pm0.150$ &0.823 &0.582 &0.039 &1.366 &  \\
		                                      &$-0.2<y<-0.1$                   &$1.959\pm0.212$ &1.228 &0.830 &0.044 &2.014 &                                                             &$1.343\pm0.135$ &0.840 &0.587 &0.041 &1.386 &  \\
		                                      &$-0.3<y<-0.2$                   &$1.804\pm0.144$ &1.146 &0.762 &0.042 &1.866  &                                                            &$1.299\pm0.122$ &0.806 &0.556 &0.039 &1.323 &  \\
		                                      &$-0.4<y<-0.3$                   &$1.854\pm0.129$ &1.189 &0.790 &0.044 &1.935   &                                                           &$1.263\pm0.091$ &0.805 &0.555 &0.040 &1.320  \\
		                                      &$-0.5<y<-0.4$                   &$1.891\pm0.118$ &1.201 &0.798 &0.045 &1.954 &                                                             &$1.238\pm0.080$ &0.798 &0.550 &0.041 &1.307  \\
	\hline

    \multirow{5}{*}{$10-20\%$}  &$-0.1<y<0~~~~~$             &$1.167\pm0.155$ &0.714 &0.464 &0.024 &1.154   &                                                           &$0.772\pm0.117$ &0.477 &0.323 &0.022 &0.778 &  \\
		                                      &$-0.2<y<-0.1$                   &$1.201\pm0.150$ &0.761 &0.486 &0.026 &1.221  &                                                            &$0.801\pm0.084$ &0.507 &0.339 &0.024 &0.822 & \\
		                                      &$-0.3<y<-0.2$                   &$1.169\pm0.114$ &0.764 &0.483 &0.028 &1.219 &                                                             &$0.830\pm0.066$ &0.533 &0.352 &0.025 &0.860 & \\
		                                      &$-0.4<y<-0.3$                   &$1.252\pm0.091$ &0.817 &0.515 &0.031 &1.301 &                                                             &$0.861\pm0.057$ &0.556 &0.367 &0.027 &0.896 &  \\
		                                      &$-0.5<y<-0.4$                   &$1.441\pm0.086$ &0.939 &0.584 &0.037 &1.486 &                                                             &$0.955\pm0.072$ &0.624 &0.408 &0.033 &0.999 &  \\
	\hline
			
    \multirow{5}{*}{$20-40\%$}  &$-0.1<y<0~~~~~$             &$0.488\pm0.064$ &0.315 &0.190 &0.009 &0.496   &                                                           &$0.319\pm0.030$ &0.203 &0.129 &0.008 &0.324 &  \\
		                                      &$-0.2<y<-0.1$                   &$0.522\pm0.057$ &0.350 &0.207 &0.011 &0.546   &                                                           &$0.345\pm0.041$ &0.228 &0.143 &0.010 &0.361 &  \\
		                                      &$-0.3<y<-0.2$                   &$0.543\pm0.071$ &0.375 &0.217 &0.013 &0.579   &                                                           &$0.379\pm0.049$ &0.256 &0.158 &0.011 &0.403 &  \\
		                                      &$-0.4<y<-0.3$                   &$0.647\pm0.048$ &0.431 &0.249 &0.015 &0.665   &                                                           &$0.438\pm0.029$ &0.289 &0.177 &0.013 &0.453 &  \\
		                                      &$-0.5<y<-0.4$                   &$0.803\pm0.080$ &0.553 &0.318 &0.022 &0.849   &                                                           &$0.553\pm0.036$ &0.369 &0.226 &0.019 &0.576 &  \\
	\hline

    \multirow{5}{*}{$40-80\%$}  &$-0.1<y<0~~~~~$             &$0.074\pm0.015$ &0.074  &0.031 &0.002 &0.103 &                                                            &$0.047\pm0.010$ &0.047 &0.023 &0.002 &0.068 &  \\
		                                      &$-0.2<y<-0.1$                   &$0.086\pm0.015$ &0.087 &0.037 &0.003 &0.121  &                                                           &$0.055\pm0.007$ &0.057 &0.027 &0.002 &0.082 &  \\
		                                      &$-0.3<y<-0.2$                   &$0.088\pm0.006$ &0.089 &0.038 &0.003 &0.124  &                                                           &$0.064\pm0.009$ &0.065 &0.031 &0.002 &0.094 &  \\
		                                      &$-0.4<y<-0.3$                   &$0.116\pm0.007$ &0.119 &0.049 &0.005 &0.163  &                                                           &$0.081\pm0.005$ &0.083 &0.038 &0.003 &0.118  &  \\
		                                      &$-0.5<y<-0.4$                   &$0.172\pm0.010$ &0.188 &0.075 &0.008 &0.255 &                                                            &$0.120\pm0.010$ &0.131 &0.060 &0.007 &0.184 &  \\

		\hline\hline
		\label{tab:dNdy-tHe3}
	\end{tabular}
\end{table*}

\begin{table*}[htbp]
	\centering
	\caption{Yield rapidity densities $dN/dy$ of $^4$He in different rapidity intervals and different centralities in Au-Au collisions at $\sqrt{s_{NN}}=3$ GeV.
Data are from Ref.~\cite{Liu:2021xfy}, and the errors denote the systematical uncertainties. The last column is the averaged deviation degree $\delta_{devi}$ of $d$, $t$, $^3$He and $^4$He.}
	\begin{tabular}{ccccccccccccccccccc}
		\toprule
	Centrality                              &Rapidity &                         &Data &                  &Theo$_{\text{total}}$ &      &Theo$_{ppnn}$ &     &Theo$_{pnd}$ &      &Theo$_{pt}$ &      &Theo$_{n^3\text{He}}$ &     &Theo$_{dd}$ & &$~~\delta_{devi}~~$      \\
      \hline
	\multirow{5}{*}{$0-10\%$}  &$-0.1<y<0~~~~~$ &          &$0.2160\pm0.0230$ &     &0.1490 &                    &0.0320 &                    &0.0192 &                   &0.0416 &                &0.0386 &                                &0.0176&    &8.3\%    \\
		                                      &$-0.2<y<-0.1$ &                &$0.2139\pm0.0185$ &     &0.1571 &                    &0.0342 &                    &0.0202 &                   &0.0439 &                &0.0406 &                                &0.0182&    &8.3\%      \\
		                                      &$-0.3<y<-0.2$ &                &$0.2058\pm0.0192$ &     &0.1513 &                    &0.0335 &                    &0.0194  &                  &0.0423 &                &0.0389 &                                &0.0172&    &9.0\%      \\
		                                      &$-0.4<y<-0.3$ &                &$0.1971\pm0.0122$ &     &0.1562 &                    &0.0346 &                    &0.0201 &                   &0.0437 &                &0.0401 &                                &0.0177&    &7.5\%      \\
		                                      &$-0.5<y<-0.4$ &                &$0.1840\pm0.0101$ &     &0.1595 &                    &0.0353 &                    &0.0205 &                   &0.0446 &                &0.0410 &                                &0.0181&    &5.8\%      \\
	\hline

    \multirow{5}{*}{$10-20\%$}  &$-0.1<y<0~~~~~$ &          &$0.1056\pm0.0181$ &     &0.0842 &                    &0.0190 &                    &0.0108 &                   &0.0236 &                &0.0215 &                                & 0.0093&    &6.3\%     \\
		                                      &$-0.2<y<-0.1$ &                &$0.1129\pm0.0205$ &     &0.0939 &                    &0.0216 &                    &0.0121 &                   &0.0263 &                &0.0238 &                                &0.0101&    &5.3\%     \\
		                                      &$-0.3<y<-0.2$ &                &$0.1220\pm0.0122$ &     &0.1004 &                    &0.0233 &                    &0.0129 &                   &0.0281 &                &0.0254 &                                &0.0107&    &7.0\%      \\
		                                      &$-0.4<y<-0.3$ &                &$0.1408\pm0.0086$ &     &0.1087 &                    &0.0254 &                    &0.0139 &                   &0.0304 &                &0.0275 &                                &0.0115&    &8.2\%      \\
		                                      &$-0.5<y<-0.4$ &                &$0.1605\pm0.0090$ &     &0.1322 &                    &0.0312 &                    &0.0169 &                   &0.0370 &                &0.0333 &                                &0.0138&    &7.6\%      \\
	\hline
			
    \multirow{5}{*}{$20-40\%$}  &$-0.1<y<0~~~~~$ &          &$0.0427\pm0.0029$ &     &0.0328 &                    &0.0080 &                    &0.0042 &                   &0.0092 &                &0.0081 &                                & 0.0033&    &7.3\%     \\
		                                      &$-0.2<y<-0.1$ &                &$0.0465\pm0.0027$ &     &0.0392  &                   &0.0097 &                    &0.0050 &                   &0.0110 &                &0.0097 &                                &0.0038&    &7.6\%      \\
		                                      &$-0.3<y<-0.2$ &                &$0.0555\pm0.0041$ &     &0.0458 &                    &0.0116 &                    &0.0059 &                   &0.0128 &                &0.0112 &                                &0.0043&    &9.8\%      \\
		                                      &$-0.4<y<-0.3$ &                &$0.0696\pm0.0087$ &     &0.0550 &                    &0.0139 &                    &0.0070 &                   &0.0154 &                &0.0135 &                                &0.0052&    &8.5\%      \\
		                                      &$-0.5<y<-0.4$ &                &$0.0967\pm0.0081$ &     &0.0785 &                    &0.0200 &                    &0.0100 &                   &0.0220 &                &0.0192 &                                &0.0073&    &9.2\%      \\
	\hline

    \multirow{5}{*}{$40-80\%$}  &$-0.1<y<0~~~~~$ &          &$0.0034\pm0.0006$ &     &0.0079 &                      &0.0026 &                    &0.0010 &                   &0.0021 &                &0.0017 &                                & 0.0005&    &55.2\% (6.2\%)     \\
		                                      &$-0.2<y<-0.1$ &                &$0.0044\pm0.0009$ &     &0.0102 &                      &0.0034 &                    &0.0013 &                   &0.0028 &                &0.0021 &                                &0.0006&    &57.3\% (6.7\%)      \\
		                                      &$-0.3<y<-0.2$ &                &$0.0059\pm0.0004$ &     &0.0116 &                      &0.0039 &                    &0.0015 &                   &0.0031 &                &0.0024 &                                &0.0007&    &48.8\% (10.2\%)      \\
		                                      &$-0.4<y<-0.3$ &                &$0.0098\pm0.0010$ &     &0.0170 &                      &0.0059 &                    &0.0021 &                   &0.0046 &                &0.0035 &                                &0.0009&    &43.3\% (12.7\%)      \\
		                                      &$-0.5<y<-0.4$ &                &$0.0159\pm0.0031$ &     &0.0318 &                      &0.0113 &                    &0.0039 &                   &0.0085 &                &0.0064 &                                &0.0017&    &53.6\% (12.8\%)      \\

		\hline\hline
		\label{tab:dNdy-He4}
	\end{tabular}
\end{table*}

To see contribution proportions of different coalescence sources of $t$, $^3$He and $^4$He in their production and depletion proportions of $d$, $t$ and $^3$He more clearly, we in this subsection study the yield rapidity densities $dN/dy$ of light nuclei.
After integrating over the $p_T$, we can get $dN/dy$.
Table \ref{tab:dNdy-d} shows our results of $d$ and Table \ref{tab:dNdy-tHe3} shows those of $t$ and $^3$He in different rapidity intervals and different centralities in Au-Au collisions at $\sqrt{s_{NN}}=3$ GeV.
Data with errors are from Ref.~\cite{Liu:2021xfy}, and the errors denote the systematical uncertainties.
Theo$_{pn}$ in the fourth column in Table \ref{tab:dNdy-d} denotes the result of $p+n$ coalescing into $d$.
Theo$_{nnp}$ and Theo$_{nd}$ in the fourth and fifth columns in Table \ref{tab:dNdy-tHe3} denote the result of $n+n+p$ coalescing into $t$ and that of $n+d$ coalescing into $t$.
Theo$_{ppn}$ and Theo$_{pd}$ in the ninth and tenth columns in Table \ref{tab:dNdy-tHe3} denote the result of $p+p+n$ coalescing into $^3$He and that of $p+d$ coalescing into $^3$He.
Theo$_{\text{dep}}$ in the fifth column of Table \ref{tab:dNdy-d} and in the sixth and eleventh columns of Table \ref{tab:dNdy-tHe3} denote the consumed $d$, $t$ and $^3$He in the nucleus coalescence process where they capture other particles to form objects with larger mass numbers.
Theo$_{\text{fin}}$ in the sixth column of Table \ref{tab:dNdy-d} and in the seventh and twelfth columns of Table \ref{tab:dNdy-tHe3} denote the final-state $d$, $t$ and $^3$He.
From Tables \ref{tab:dNdy-d} and \ref{tab:dNdy-tHe3}, one can see that our results Theo$_{\text{fin}}$ agree well with the experimental data in 0-10\%, 10-20\% and 20-40\% centralities.
But in the peripheral 40-80\% centrality, our Theo$_{\text{fin}}$ of $d$ underestimates the data and Theo$_{\text{fin}}$ of $t$ and $^3$He overestimates the data;
our results only including nucleon coalescence Theo$_{pn}$, Theo$_{nnp}$ and Theo$_{ppn}$ can describe the corresponding data much better.
This further indicates that nucleon coalescence is the dominant production for light nuclei in peripheral $40-80$\% collisions, and other coalescence channels involving nucleon$+$nucleus and nucleus$+$nucleus may not occur.

Table \ref{tab:dNdy-He4} shows results of $^4$He in different rapidity intervals and different centralities in Au-Au collisions at $\sqrt{s_{NN}}=3$ GeV.
Data with errors are from Ref.~\cite{Liu:2021xfy}, and the errors denote the systematical uncertainties.
Theo$_{ppnn}$, Theo$_{pnd}$, Theo$_{pt}$, Theo$_{n^3\text{He}}$ and Theo$_{dd}$ in the fifth, sixth, seventh, eighth and ninth columns denote the results of $p+p+n+n$, $p+n+d$, $p+t$, $n+^3$He and $d+d$ coalescing into $^4$He, respectively.
Theo$_{\text{total}}$ in fourth column denote total results  including all five coalescence sources for $^4$He.
Theo$_{\text{total}}$ in 0-10\%, 10-20\% and 20-40\% centralities and Theo$_{ppnn}$ in the peripheral 40-80\% centrality give about $20\% \sim 30$\% underestimations of the central values of the experimental data.
This may be due to that we do not consider decay contributions from the excited states of $^4$He.
If decay properties of these excited states become clear and these contributions are included in the future, theoretical results will approach to the data better.
We employ the averaged deviation degree $\delta_{devi}$ to quantitatively characterize the deviation extent of our theoretical results from the data, and it is defined as
{\setlength\arraycolsep{0pt}
\begin{eqnarray}
 \delta_{devi} = \frac{1}{4} \sum\limits_{j=d,t,^3\text{He},^4\text{He}} \left| \frac{\text{Theory}_{j}-\text{Data}_{j}}{\text{Data}_{j}} \right| .   \label{eq:deviation}  
\end{eqnarray} }%
Values of $\delta_{devi}$ calculated with Theo$_{\text{fin}}$ for $d$, $t$, $^3$He and Theo$_{\text{total}}$ for $^4$He are put in the last column in Table \ref{tab:dNdy-He4}, and those in the parentheses for the $40-80\%$ centrality are calculated with the results only including nucleon coalescence.

Our theoretical results in Tables \ref{tab:dNdy-d}, \ref{tab:dNdy-tHe3} and \ref{tab:dNdy-He4} clearly show contribution proportions of different production sources for $d$, $t$, $^3$He and $^4$He in their production in $0-10$\%, $10-20$\% and $20-40$\% centralities.
The proportion of nucleon coalescence and that of nucleon$+d$ coalescence in $t$ and $^3$He production take about 60\% and 40\%, respectively.
The proportion of nucleon coalescence and those of $p+n+d$ coalescence, $p+t$ coalescence, $n+^3$He coalescence and $d+d$ coalescence in $^4$He production take about 20\%, 15\%, 30\%, 25\% and 10\%, respectively.
Tables \ref{tab:dNdy-d} and \ref{tab:dNdy-tHe3} also show that the depletion of $d$ takes about $7\%\sim9\%$ while the depletions of $t$ and $^3$He are both less than 3\%. 
These results tell us that besides nucleon coalescence, other particle coalescences, e.g., composite particles of less mass numbers coalescing into light nuclei of larger mass numbers or composite particles capturing nucleons to recombine into heavier light nuclei, also play important roles in light nuclei production in central and semi-central collisions at relatively low collision energies.
This provides a new possible window to cognize the underestimations of the yield densities of light nuclei in some specific models only including nucleon coalescence such as in Ref.~\cite{Bratkovskaya:2022vqi}.

%------------------------------------------------------------------------------------------------------------------------------ light nuclei <pT>
\subsection{Averaged transverse momenta of light nuclei}

\begin{table*}[htbp]
	\centering
	\caption{Averaged transverse momenta $\langle p_T \rangle$ of $d$, $t$ and $^3$He in different rapidity intervals and different centralities in Au-Au collisions at $\sqrt{s_{NN}}=3$ GeV.
Data are from Ref.~\cite{Liu:2021xfy}, and the errors denote the systematical uncertainties.}
	\begin{tabular}{ccccccccccccccc}
		\toprule
	\multirow{2}{*}{Centrality}  &\multirow{2}{*}{Rapidity}  &$d$ &                     &      &       &$t$                 &                    &                   &     &       &$^3$He         &                    &                   &                     \\
                                                                                                 \cline{3-4}\cline{6-9}\cline{11-14}
		                                      &          &Data &$\langle p_T\rangle_{\text{fin}}$ &   &Data &$\langle p_T\rangle_{\text{fin}}$ &$\langle p_T\rangle_{nnp}$ &$\langle p_T\rangle_{nd}$  &     &Data &$\langle p_T\rangle_{\text{fin}}$ &$\langle p_T\rangle_{ppn}$ &$\langle p_T\rangle_{pd}$ &             \\
      \hline
	\multirow{5}{*}{$0-10\%$}  &$-0.1<y<0~~~~~$             &$1.048\pm0.033$ &1.033 &       &$1.363\pm0.044$ &1.343 &1.347 &1.337 &                           &$1.412\pm0.044$ &1.340 &1.344 &1.335 &  \\
		                                      &$-0.2<y<-0.1$                   &$1.049\pm0.032$ &1.028 &       &$1.350\pm0.047$ &1.338 &1.342 &1.332 &                            &$1.405\pm0.041$ &1.335 &1.338 &1.330 &  \\
		                                      &$-0.3<y<-0.2$                   &$1.036\pm0.026$ &1.015 &       &$1.320\pm0.037$ &1.318 &1.322 &1.312 &                            &$1.384\pm0.045$ &1.315 &1.318 &1.310 &  \\
		                                      &$-0.4<y<-0.3$                   &$1.019\pm0.031$ &1.004 &       &$1.291\pm0.030$ &1.308 &1.312 &1.302 &                            &$1.358\pm0.034$ &1.305 &1.309 &1.300 &  \\
		                                      &$-0.5<y<-0.4$                   &$0.987\pm0.024$ &0.976 &       &$1.242\pm0.020$ &1.274 &1.277 &1.268 &                            &$1.308\pm0.024$ &1.271 &1.274 &1.266 &  \\
	\hline

    \multirow{5}{*}{$10-20\%$}  &$-0.1<y<0~~~~~$             &$0.996\pm0.042$   &0.965 &       &$1.256\pm0.043$ &1.242 &1.246 &1.236 &                           &$1.306\pm0.052$ &1.239 &1.243 &1.234 &  \\
		                                      &$-0.2<y<-0.1$                   &$0.992\pm0.045$   &0.962 &       &$1.239\pm0.035$ &1.238 &1.242 &1.232 &                            &$1.297\pm0.039$ &1.235 &1.238 &1.230 &  \\
		                                      &$-0.3<y<-0.2$                   &$0.974\pm0.024$   &0.941 &        &$1.217\pm0.044$ &1.209 &1.213 &1.203 &                            &$1.275\pm0.030$ &1.206 &1.209 &1.201 &  \\
		                                      &$-0.4<y<-0.3$                   &$0.956\pm0.025$   &0.935 &        &$1.200\pm0.030$ &1.204 &1.208 &1.199 &                            &$1.261\pm0.025$ &1.202 &1.205 &1.197 &  \\
		                                      &$-0.5<y<-0.4$                   &$0.924\pm0.032$   &0.919 &        &$1.164\pm0.014$ &1.188 &1.191 &1.182 &                            &$1.217\pm0.032$ &1.185 &1.188 &1.180 &  \\
	\hline
			
    \multirow{5}{*}{$20-40\%$}  &$-0.1<y<0~~~~~$             &$0.908\pm0.034$  &0.893 &       &$1.136\pm0.053$ &1.137 &1.140 &1.131 &                            &$1.183\pm0.050$ &1.134 &1.137 &1.129 &  \\
		                                      &$-0.2<y<-0.1$                   &$0.898\pm0.042$  &0.888 &       &$1.122\pm0.043$ &1.129 &1.132 &1.123 &                            &$1.171\pm0.053$ &1.126 &1.129 &1.121 &  \\
		                                      &$-0.3<y<-0.2$                   &$0.880\pm0.034$  &0.872 &       &$1.093\pm0.051$ &1.109 &1.112 &1.103 &                            &$1.153\pm0.057$ &1.106 &1.109 &1.101 &  \\
		                                      &$-0.4<y<-0.3$                   &$0.869\pm0.031$  &0.863 &       &$1.067\pm0.027$ &1.097 &1.100 &1.092 &                            &$1.115\pm0.021$ &1.094 &1.097 &1.090 &  \\
		                                      &$-0.5<y<-0.4$                   &$0.833\pm0.019$  &0.834 &       &$1.039\pm0.041$ &1.061 &1.064 &1.056 &                            &$1.063\pm0.020$ &1.059 &1.061 &1.055 &  \\
	\hline

    \multirow{5}{*}{$40-80\%$}  &$-0.1<y<0~~~~~$             &$0.779\pm0.023$  &0.779 &       &$0.925\pm0.019$ &0.971  &0.974 &0.966 &                           &$0.978\pm0.046$ &0.968 &0.970 &0.964 &  \\
		                                      &$-0.2<y<-0.1$                   &$0.774\pm0.001$  &0.767 &       &$0.911\pm0.028$ &0.952 &0.954 &0.946 &                            &$0.947\pm0.026$ &0.949 &0.951 &0.945 &  \\
		                                      &$-0.3<y<-0.2$                   &$0.761\pm0.023$  &0.748 &       &$0.900\pm0.011$ &0.927 &0.929 &0.922 &                            &$0.931\pm0.032$ &0.925 &0.926 &0.921 &  \\
		                                      &$-0.4<y<-0.3$                   &$0.736\pm0.001$  &0.739 &       &$0878\pm0.011$ &0.916 &0.918 &0.911 &                             &$0.899\pm0.004$ &0.914 &0.915 &0.910 &  \\
		                                      &$-0.5<y<-0.4$                   &$0.706\pm0.013$  &0.719 &       &$0.833\pm0.009$ &0.895 &0.897 &0.890 &                            &$0.854\pm0.025$ &0.893 &0.894 &0.889 &  \\

		\hline\hline
		\label{tab:avepT-dtHe3}
	\end{tabular}
\end{table*}

\begin{table*}[htbp]
	\centering
	\caption{Averaged transverse momenta $\langle p_T \rangle$ of $^4$He in different rapidity intervals and different centralities in Au-Au collisions at $\sqrt{s_{NN}}=3$ GeV.
Data are from Ref.~\cite{Liu:2021xfy}, and the errors denote the systematical uncertainties.}
	\begin{tabular}{cccccccccccccccccc}
		\toprule
	Centrality                              &Rapidity &                         &Data &     &$\langle p_T\rangle_{\text{total}}$ &  &$\langle p_T\rangle_{ppnn}$ &  &$\langle p_T\rangle_{pnd}$ &      &$\langle p_T\rangle_{pt}$ &      &$\langle p_T\rangle_{n^3\text{He}}$ &     &$\langle p_T\rangle_{dd}$ &      \\
      \hline
	\multirow{5}{*}{$0-10\%$}  &$-0.1<y<0~~~~~$ &          &$1.591\pm0.048$ &     &1.621 &                    &1.631 &                    &1.620 &                   &1.621 &                &1.618 &                                &1.609     \\
		                                      &$-0.2<y<-0.1$ &                &$1.566\pm0.041$ &     &1.615 &                    &1.625 &                    &1.614 &                   &1.615 &                &1.612 &                                &1.603      \\
		                                      &$-0.3<y<-0.2$ &                &$1.535\pm0.046$ &     &1.588 &                    &1.598 &                    &1.587  &                  &1.588 &                &1.585 &                                &1.576      \\
		                                      &$-0.4<y<-0.3$ &                &$1.508\pm0.022$ &     &1.582 &                    &1.591 &                    &1.581 &                   &1.582 &                &1.579 &                                &1.570      \\
		                                      &$-0.5<y<-0.4$ &                &$1.483\pm0.003$ &     &1.542 &                    &1.551 &                    &1.541 &                   &1.542 &                &1.539 &                                &1.531      \\
	\hline

    \multirow{5}{*}{$10-20\%$}  &$-0.1<y<0~~~~~$ &          &$1.496\pm0.081$ &     &1.487 &                    &1.496 &                    &1.486 &                   &1.486 &                &1.484 &                                &1.475    \\
		                                      &$-0.2<y<-0.1$ &                &$1.487\pm0.080$ &     &1.481 &                    &1.491 &                    &1.480 &                   &1.481 &                &1.478 &                                &1.469      \\
		                                      &$-0.3<y<-0.2$ &                &$1.446\pm0.046$ &     &1.444 &                    &1.454 &                    &1.443 &                   &1.444 &                &1.442 &                                &1.433      \\
		                                      &$-0.4<y<-0.3$ &                &$1.397\pm0.020$ &     &1.442 &                    &1.451 &                    &1.441 &                   &1.441 &                &1.439 &                                &1.430      \\
		                                      &$-0.5<y<-0.4$ &                &$1.363\pm0.006$ &     &1.426 &                    &1.435 &                    &1.425 &                   &1.426 &                &1.423 &                                &1.415      \\
	\hline
			
    \multirow{5}{*}{$20-40\%$}  &$-0.1<y<0~~~~~$ &          &$1.316\pm0.036$ &     &1.348 &                    &1.357 &                    &1.347 &                   &1.348 &                &1.345 &                                &1.337     \\
		                                      &$-0.2<y<-0.1$ &                &$1.296\pm0.024$ &     &1.337 &                    &1.346 &                    &1.336 &                   &1.337 &                &1.334 &                                &1.326      \\
		                                      &$-0.3<y<-0.2$ &                &$1.262\pm0.006$ &     &1.314 &                    &1.322 &                    &1.312 &                   &1.313 &                &1.311 &                                &1.302      \\
		                                      &$-0.4<y<-0.3$ &                &$1.227\pm0.058$ &     &1.300 &                    &1.308 &                    &1.299 &                   &1.299 &                &1.297 &                                &1.289      \\
		                                      &$-0.5<y<-0.4$ &                &$1.173\pm0.038$ &     &1.259 &                    &1.266 &                    &1.257 &                   &1.258 &                &1.256 &                                &1.248      \\
	\hline

    \multirow{5}{*}{$40-80\%$}  &$-0.1<y<0~~~~~$ &          &$1.139\pm0.048$ &     &1.135 &                      &1.142 &                    &1.133 &                   &1.133 &                &1.130 &                                &1.123   \\
		                                      &$-0.2<y<-0.1$ &                &$1.095\pm0.043$ &     &1.109 &                      &1.115 &                    &1.107 &                   &1.107 &                &1.104 &                                &1.098   \\
		                                      &$-0.3<y<-0.2$ &                &$1.062\pm0.005$ &     &1.075 &                      &1.080 &                    &1.072 &                   &1.072 &                &1.070 &                                &1.064   \\
		                                      &$-0.4<y<-0.3$ &                &$1.005\pm0.026$ &     &1.066 &                      &1.072 &                    &1.064 &                   &1.064 &                &1.062 &                                &1.056  \\
		                                      &$-0.5<y<-0.4$ &                &$0.972\pm0.072$ &     &1.044  &                     &1.050 &                    &1.043 &                   &1.043 &                &1.041 &                                &1.035   \\

		\hline\hline
		\label{tab:avepT-He4}
	\end{tabular}
\end{table*}

The averaged transverse momenta of different light nuclei reflect the collective motion and bulk properties of the hadronic matter at kinetic freezeout.
In this subsection we study the averaged transverse momenta $\langle p_T \rangle$ of $d$, $t$, $^3$He and $^4$He in rapidity intervals $-0.1<y<0$, $-0.2<y<-0.1$, $-0.3<y<-0.2$, $-0.4<y<-0.3$, $-0.5<y<-0.4$ in Au-Au collisions at $\sqrt{s_{NN}}=3$ GeV in centralities $0-10$\%, $10-20$\%, $20-40$\%, $40-80$\%, respectively.
Table \ref{tab:avepT-dtHe3} and Table \ref{tab:avepT-He4} show the results.
Data with errors are from Ref.~\cite{Liu:2021xfy}, and the errors denote the systematical uncertainties.
The $\langle p_T \rangle_{\text{fin}}$ in the fourth, sixth and tenth columns in Table \ref{tab:avepT-dtHe3} denotes our theoretical results for final-state $d$, $t$, $^3$He, respectively, 
and $\langle p_T \rangle_{\text{total}}$ in the fourth column in Table \ref{tab:avepT-He4} denotes total results  including all five coalescence sources for $^4$He.
$\langle p_T \rangle_{nnp}$ and $\langle p_T \rangle_{nd}$ in the seventh and eighth columns in Table \ref{tab:avepT-dtHe3} denote the result of $n+n+p$ coalescing into $t$ and that of $n+d$ coalescing into $t$.
$\langle p_T \rangle_{ppn}$ and $\langle p_T \rangle_{pd}$ in the eleventh and twelfth columns in Table \ref{tab:avepT-dtHe3} denote the result of $p+p+n$ coalescing into $^3$He and that of $p+d$ coalescing into $^3$He.
$\langle p_T \rangle_{ppnn}$, $\langle p_T \rangle_{pnd}$, $\langle p_T \rangle_{pt}$, $\langle p_T \rangle_{n^3\text{He}}$ and $\langle p_T \rangle_{dd}$ in the fifth, sixth, seventh, eighth and ninth columns in Table \ref{tab:avepT-He4} denote the results of $p+p+n+n$, $p+n+d$, $p+t$, $n+^3$He and $d+d$ coalescing into $^4$He, respectively.

Table \ref{tab:avepT-dtHe3} and Table \ref{tab:avepT-He4} show that for $t$, $^3$He and $^4$He, the calculated $\langle p_T \rangle$ from different coalescence sources are almost the same.
This is very different from $dN/dy$.
Our theoretical results agree with the data, and the deviations are less than 10\%.
$\langle p_T \rangle$ of $d$, $t$, $^3$He and $^4$He decreases gradually as the increasing rapidity from central to peripheral collisions.
This further indicates the stronger transverse collective motion at midrapidity area in more central collisions.

At the end of Sec.~\ref{results}, we want to state that our results show the coalescence mechanism still works in describing light nuclei production in Au-Au collisions at $\sqrt{s_{NN}}=3$ GeV.
Compared to those at high RHIC and LHC energies in our previous works~\cite{Wang:2020zaw,Zhao:2022xkz}, relativistic heavy ion collisions at lower collision energies have some new characteristics in light nuclei production, e.g., isospin asymmetry from the colliding nuclei and the non-negligible nucleus$+$nucleon/nucleus coalescence.

%----------------------------------------------------------------------------------------------------------------------------------------------------------------- summary
\section{summary}    \label{summary}

In the coalescence mechanism, we studied different coalescence sources of the production of various species of light nuclei in relativistic heavy ion collisions. 
We firstly extended the coalescence model to include two bodies, three bodies, and four nucleons coalescing into light nuclei, respectively. 
We used the assumption of the coordinate-momentum factorization of joint hadronic distributions.
We adopted gaussian forms for the relative coordinate distributions.
Based on these simplifications, we obtained analytic formulas of momentum distributions of light nuclei formed from different production sources which coalesced by different hadrons.

We then applied the extended coalescence model to Au-Au collisions at $\sqrt{s_{NN}}=3$ GeV to simultaneously investigate the $p_T$ spectra of the $d$, $t$, $^3$He and $^4$He in different rapidity intervals at midrapidity area from central to peripheral collisions.
We presented the $p_T$ dependence of different coalescence sources for $d$, $t$, $^3$He and $^4$He.
We also studied yield rapidity densities $dN/dy$ and averaged transverse momenta $\langle p_T \rangle$ of $d$, $t$, $^3$He and $^4$He.
We gave proportions of yield densities from different coalescence sources for $t$, $^3$He and $^4$He in their production and those of depletions for $d$, $t$ and $^3$He.
We found yield densities from different coalescence sources for a specific kind of light nuclei were very different, but averaged transverse momenta were almost unchanged.

Our results showed that (1) results of $p+n$ coalescence minus those depleted in nucleus coalescence reproduced the available data of $d$ well in central and semi-central collisions and the data in peripheral collisions favored $p+n$ coalescence;
(2) the nucleon coalescence plus nucleon$+d$ coalescence reproduced the available data of $t$ and $^3$He in central and semi-central collisions (their depletions in forming $^4$He are less than 3\%) and the data in peripheral collisions favored only nucleon coalescence;
(3) the nucleon coalescence plus nucleon$+$nucleus coalescence and nucleus$+$nucleus coalescence described the available data of $^4$He in central and semi-central collisions and the data in peripheral collisions favored only $p+p+n+n$ coalescence.

%---------------------------------------------------------------------------------------------------------------------------------------------------------------------------- 
\section*{Acknowledgements}

We thank Prof. Xiao-Feng Luo for helpful discussions and thank the STAR collaboration for providing us the preliminary data of $p_T$ spectra of $d$, $t$, $^3$He and $^4$He in Au-Au collisions at $\sqrt{s_{NN}}=3$ GeV.
This work was supported in part by the National Natural Science Foundation of China under Grants No. 12175115 and No. 12375074, the Natural Science Foundation of Shandong Province, China, under Grants No. ZR2020MA097
and Higher Educational Youth Innovation Science and Technology Program of Shandong Province under Grants No. 2020KJJ004 and No. 2019KJJ010.

\bibliographystyle{apsrev4-1}
\bibliography{ref}

\end{document}